\pdfoutput=1
\documentclass[a4paper,11pt]{article}

\usepackage{lineno}
\usepackage{jheppub}

\usepackage[utf8]{inputenc}
\usepackage{amsmath}
\usepackage{amsthm}
\usepackage{amssymb}
\usepackage{booktabs} 
\usepackage{enumitem}   
\usepackage[english]{babel}
\usepackage{url}
\usepackage{mathtools}
\usepackage{bbold}
\usepackage{slashed}
\usepackage{multirow}
\usepackage{lipsum}
\usepackage{xcolor}
\usepackage[dvipsnames]{xcolor}
\usepackage[table]{xcolor}
\usepackage{float}
\usepackage{revsymb}
\usepackage{braket}
\usepackage{siunitx}
\usepackage{xfrac}
\usepackage{physics}
\usepackage[subpreambles=true]{standalone}
\usepackage{comment}

\allowdisplaybreaks

\begin{document}


\title{Minimal spontaneous baryogenesis from flavor}

\author[a,b]{Martin A. Mojahed,}
\author[a,b]{Alfredo Stanzione,}

\affiliation[a]{INFN Sezione di Roma, Piazzale Aldo Moro 2, I-00185 Rome, Italy}
\affiliation[b]{Università degli Studi di Roma La Sapienza, Piazzale Aldo Moro 5, 00185, Rome, Italy}


\abstract{
In its minimal realization, the Froggatt–Nielsen (FN) mechanism addresses the Standard Model (SM) flavor puzzle via the spontaneous breaking of a horizontal $U(1)_H$ symmetry. If $U(1)_H$ is global and broken before the end of inflation, it gives rise to a pseudo-Nambu-Goldstone boson (axion-like particle) with a homogeneous initial misalignment, whose subsequent coherent evolution sources spontaneous $CPT$ violation in the SM sector. We show that in the presence of $B-L$-violating processes—parametrized here model-independently via the dimension-five Weinberg operator responsible for light neutrino masses—the minimal FN setup can generate the baryon asymmetry of the Universe (BAU), linking the origin of flavor directly to the BAU and neutrino masses. Solving a complete set of evolution equations for SM asymmetries, we demonstrate that this mechanism can readily yield a baryon asymmetry comparable to the observed value, or even exceeding it by several orders of magnitude. Accounting for baryonic isocurvature and entropy dilution from axion decays, we find that the observed BAU is successfully reproduced for axion oscillation temperatures above $10^{11}$~GeV and axion decay constants ranging from $10^{13\cdots 14}$~GeV to above $10^{17}$~GeV. 
Finally, while we focus on a global $U(1)_H$ symmetry and axion motion from standard misalignment, our framework is readily extendable to other global horizontal symmetries and arbitrary axion motion.
}


\maketitle


\section{Introduction}
The mass spectrum of quarks spans five orders of magnitude, and the mass hierarchy in the leptonic sector spans at least ten orders of magnitude, depending on the absolute neutrino mass scale. Moreover, the quark sector features a hierarchical misalignment between the gauge and mass eigenstates, as encoded in the Cabibbo-Kobayashi–Maskawa (CKM) matrix~\cite{Cabibbo:1963yz,Kobayashi:1973fv}, while leptons have three sizable mixing angles in the Pontecorvo–Maki-Nakagawa-Sakata (PMNS) matrix~\cite{Pontecorvo:1957qd,Maki:1962mu}. While this dramatic variation in masses and mixings is technically natural, it invites a dynamical explanation. Indeed, a primary goal of flavor physics is to understand the appearance of large hierarchies in the masses and mixing angles of the Standard Model (SM) fermions. 

A popular solution to the above SM flavor puzzle in the context of quantum-field theory in four dimensions is provided by the Froggatt-Nielsen (FN) models~\cite{Froggatt:1978nt} (and their variations~\cite{Leurer:1992wg,Leurer:1993gy}, see also Refs.~\cite{Nir:1996am,Dudas:1996fe,Irges:1998ax,Sato:2000ff,Sato:2000kj,Dreiner:2003hw,Babu:2003zz,Bauer:2016rxs,Feruglio:2019ybq,Nishimura:2020nre,Smolkovic:2019jow,Fedele:2020fvh,Aloni:2021wzk,Greljo:2023bix,Asadi:2023ucx,Cornella:2023zme,Ibe:2024cvi,Loisa:2024xuk,Cornella:2024jaw,Greljo:2024evt}). In the most standard FN models, SM fermions are charged under a new abelian horizontal symmetry, $U(1)_H$, which is eventually spontaneously broken by a scalar field, known as the flavon. The new symmetry forbids generic SM Yukawa couplings at tree level, and effective Yukawa couplings are instead generated by non-renormalizable operators involving
the (chiral) SM fermions, the SM Higgs, and the flavon field. A standard UV perspective on the origin of these higher-dimensional operators is that they could be generated through a chain of heavy vector-like matter that also carries $U(1)_H$ charge~\cite{Froggatt:1978nt}. The dimensionality of these operators is dictated by the charges of the SM fermions under $U(1)_H$. As a result, the magnitude of the effective Yukawa couplings becomes exponentially sensitive to the FN charges of the corresponding fermions. Consequently, order-one differences in charge assignments under $U(1)_H$ generate large hierarchies in masses and mixing angles. 

Two further outstanding puzzles beyond the SM are the origin of the observed baryon asymmetry of the Universe (BAU)~\cite{Planck:2018vyg,ParticleDataGroup:2024cfk} and the origin and smallness of (active) neutrino masses. Leptogenesis (LG)~\cite{Fukugita:1986hr} provides a compelling framework for addressing both questions simultaneously: it dynamically generates the BAU in the early Universe while accounting for active neutrino masses through the type-I seesaw mechanism with at least two right-handed neutrinos (RHNs)~\cite{Minkowski:1977sc,Yanagida:1979as,Yanagida:1980xy,Gell-Mann:1979vob,Mohapatra:1979ia,Schechter:1980gr}. In the standard high-scale realization, where the RHN masses lie well above the electroweak scale, LG proceeds through CP-violating, out-of-equilibrium decays of RHNs. The lepton asymmetry that is generated through these decays is then partially converted into a baryon asymmetry by electroweak sphaleron processes, which violate baryon plus lepton number, $B+L$~\cite{Kuzmin:1985mm}. This conventional mechanism can be ineffective, however, if the RHN sector does not entail sufficient CP violation, or if the RHNs are so heavy that they are never thermally populated after inflation. The latter case represents a particularly challenging regime for LG and is the scenario studied in this work.

In particular, we will show that a spontaneously broken global horizontal symmetry $U(1)_H$ along with a source of lepton-number violation (LNV) can connect the origins of flavor, neutrino masses, and the BAU. The abelian pseudo--Goldstone boson, hereafter referred to as the axion, will allow a realization of spontaneous LG ~\cite{Cohen:1987vi,Cohen:1988kt,Chiba:2003vp,Takahashi:2003db,Kusenko:2014uta,Ibe:2015nfa,Takahashi:2015waa,Bae:2018mlv,Co:2019wyp,Domcke:2020kcp,Co:2020xlh,Co:2020jtv,Foster:2022ajl,Chao:2023ojl,Chun:2023eqc,Datta:2024xhg,Barnes:2024jap,Peng:2025sri,Chun:2025abp,Takahashi:2026ngu,Chun:2026jgn}. This LG mechanism does not rely on dynamical RHNs, \textit{any} source of high-scale LNV is sufficient. Indeed, if RHNs are heavy compared to the maximum temperature attained in the early Universe, they can be integrated out. Their effects are then encoded in the dimension-five Weinberg operator~\cite{Weinberg:1979sa} throughout the thermal history of the Universe, and it is this operator that forms the basis of our LG analysis. From a bottom-up perspective, however, the Weinberg operator need not originate from RHNs; rather, it may arise as the low-energy remnant of an alternative ultraviolet (UV) source of LNV. Consequently, assuming a non-vanishing Weinberg operator allows us to capture a significantly broader class of scenarios featuring high-scale LNV than those encompassed by the standard type-I seesaw framework.

The remainder of this paper is organized as follows. In Section~\ref{sec:2}, we briefly review aspects of the FN mechanism and spontaneous baryogenesis most relevant for our analysis. We then describe a viable cosmological history in which our framework can operate, compare our setup with related proposals in the literature, and discuss minimal modifications that could allow it to mitigate the Higgs naturalness problem, provide a DM candidate, and solve the strong CP problem. In Section~\ref{sec:3}, we present the relevant evolution equations for SM asymmetries, which we solve numerically for selected benchmark examples. We also discuss how processes in the primordial plasma influence the generation of the BAU in our framework. Section~\ref{sec:4} examines the cosmological constraints on our scenario, and identifies regions in axion-parameter space where the mechanism operates successfully. We summarize our results and discuss future directions in Section~\ref{sec:conclusion}. Some details about the consistency of our analysis can be found in Appendix~\ref{appendix:FN}.

\section{Connecting the origin of flavor structures with baryogenesis}
\label{sec:2}

The purpose of this section is to outline how the origin of flavor structure in the SM can be unified with the origin of the BAU. To this end, we start by reviewing some aspects of the most standard FN construction in Section~\ref{sec:FN}, followed by elements of spontaneous LG in Section~\ref{sec:SBG}. In Section~\ref{sec:cosmohistory}, we present a minimal cosmological history in which the unified mechanism to explain the origin of flavor and the BAU can be successfully realized. Finally, in Section~\ref{sec:comparison} we comment on similarities and differences between our proposal and some previous frameworks in the literature. Along the way, we comment on possible extensions of the minimal framework considered here.

\subsection{A brief review of the Froggatt–Nielsen mechanism}
\label{sec:FN}
The SM contains a large number of free parameters in its Yukawa sector. After electroweak symmetry breaking, these parameters determine the charged-fermion mass spectrum and the CKM mixing matrix, while extensions that account for neutrino masses introduce the PMNS matrix as an additional source of flavor structure. The experimentally inferred values of these parameters are highly non-generic, giving rise to the so-called \emph{flavor puzzle}: charged-fermion masses span several orders of magnitude, the CKM matrix is approximately diagonal with off-diagonal elements organized in powers of $\lambda \simeq 0.2$, and the PMNS matrix, in contrast, shows no comparable hierarchy and instead exhibits a largely anarchic structure. Although these features can be accommodated within the SM, their origins lack a dynamical explanation, strongly motivating the search for deeper underlying mechanisms.

The FN mechanism provides a well-established framework for addressing the origin of flavor hierarchies. In its most standard formulation, these hierarchies arise from a horizontal abelian $U(1)_H$ symmetry under which SM fermions carry generation-dependent charges, while the Higgs field $H$ is taken to be neutral. In this setup, the renormalizable Yukawa couplings $\bar{\psi}_{L,i} H \psi_{R,j}$ are forbidden whenever the $U(1)_H$ charges of $\psi_{L,i}$ and $\psi_{R,j}$ do not match. In addition to the $U(1)_H$ symmetry, the framework requires two sets of new degrees of freedom (DoF): 
\begin{enumerate}
    \item A complex scalar field $\phi$, the \emph{flavon}, carrying $U(1)_H$ charge $X_\phi = 1$, whose vacuum expectation value (vev), $\langle \phi \rangle=v_\phi/\sqrt{2}$, spontaneously breaks the flavor symmetry.
    \item A set of heavy vector-like fermions with mass scale $M > v_\phi$, which act as messengers between the flavon and the SM fields.
\end{enumerate}
Upon integrating out these heavy states, one generates at tree level higher-dimensional operators of the form
\begin{equation}
    \mathcal{L}_{\text{Y}} \supset \sum_{i,j} r_{ij} \left( \frac{\phi^{(\dagger)}}{M} \right)^{n_{ij}} \bar{\psi}_{L,i} H \psi_{R,j} + \text{h.c.},
    \label{eq:fn_lagrangian}
\end{equation}
with $r_{ij} = \mathcal{O}(1)$ and exponents $n_{ij} = |X_{\psi_{L,i}} - X_{\psi_{R,j}}|$ fixed by $U(1)_H$ selection rules. The sign of the argument inside the absolute value determines whether $\phi$ or $\phi^\dagger$ enters the operator. When the flavon acquires a nonzero vev, these operators generate effective Yukawa terms in which every entry is suppressed by a power of the small parameter $\epsilon \equiv v_\phi/M$. In this way, the structure of the SM Yukawa matrices is dynamically generated from charge differences alone, while the $r_{ij}$ remain natural $\mathcal{O}(1)$ numbers. Diagonalizing the resulting mass matrices, the physical masses and the entries of the CKM and PMNS matrices inherit a simple power-law dependence on $\epsilon$. In particular, for the general charge assignment shown in Tab.~\ref{tab:wrinkles}, fixing $\epsilon = \lambda \simeq 0.2$ makes the matching between charges especially transparent~\cite{Asadi:2023ucx}:
\begin{align}
    \label{eq:VCKM}
    V_{ij} \sim \epsilon^{|X_{Q_i}-X_{Q_j}|}, &\qquad 
    U_{ij} \sim \epsilon^{|X_{L_i}-X_{L_j}|}, \qquad \\
    m_i^{u} \sim v_H\epsilon^{|X_{Q_i}-X_{u_i}|}, \qquad
    m_i^{d} \sim &v_H\epsilon^{|X_{Q_i}-X_{d_i}|}, \qquad
    m_i^{\ell} \sim v_H\epsilon^{|X_{L_i}-X_{e_i}|}.
\end{align}

The FN mechanism only fixes the ratio $\epsilon = v_\phi / M$, while $v_\phi$ and $M$ are individually unconstrained: any choice that reproduces the required value of $\epsilon$ leads to the same hierarchical structure of the Yukawa matrices. In principle, the absolute scale can be constrained by studying, for example, tree-level exchange of the radial flavon mode, which mediates flavor-violating processes and can be confronted with the bounds from flavor physics. In this work, however, we are interested in scenarios where $v_\phi$ and $M$ lie close to the GUT scale, so that any such low-energy contribution is suppressed well below current experimental sensitivities and plays no role in the phenomenology we discuss.
\begin{table}[h]
\centering
\renewcommand{\arraystretch}{1.}
\rowcolors{2}{white}{Orange!8}
\begin{tabular}{l c c c}
\toprule
 & \textbf{Gen.\ 1} & \textbf{Gen.\ 2} & \textbf{Gen.\ 3} \\
\midrule
$Q$        & $-q_0 - 3X$        & $-q_0 - 2X$  & $-q_0$       \\
$u$  & $-q_0 - 3X \mp 7$   & $-q_0 + X$    & $-q_0$        \\
$d$  & $-q_0 - 3X \mp 6$   & $-q_0 + 3X$   & $-q_0 + 2X$   \\
$L$        & $l_0 + Y$          & $l_0$        & $l_0$        \\
$e$  & $l_0 + Y \mp 8$   & $+l_0 - 5Y$  & $+l_0 - 3Y$  \\
\bottomrule
\end{tabular}
\caption{Generic Froggatt-Nielsen charge assignments for fixed $\epsilon=0.2$ compatible with SM masses and mixing angles. In supersymmetric theories, holomorphy of the superpotential requires $X=-Y=-1$ and picks the minus sign for right-handed fermions in the first generation. Table adapted from Ref.~\cite{Asadi:2023ucx}.}
\label{tab:wrinkles}
\end{table}

Upon spontaneous $U(1)_H$ breaking, we parameterize the flavon around its minimum as $\phi = \frac{1}{\sqrt{2}}(v_\phi + \rho) e^{i a / v_\phi}$, where $\rho$ and $a$ denote the radial and angular (axion) modes, respectively. Inserting this into Eq.~\eqref{eq:fn_lagrangian} yields the general Lagrangian for the SM-Higgs-flavon sector,
\begin{align}
    \label{eq:FNL}
    \mathcal{L}\supset & +r_{ij}^u\Bar{Q}_i\Tilde{H}u_j\epsilon^{n_{ij}^u}e^{\pm ian_{ij}^u/v_\phi}+r_{ij}^d\Bar{Q}_iHd_j\epsilon^{n_{ij}^d}e^{\pm ian_{ij}^d/v_\phi}\nonumber \\&+r_{ij}^e\Bar{L}_iHe_j\epsilon^{n_{ij}^e}e^{\pm ian_{ij}^e/v_\phi}+\frac{c_{ij}\,\epsilon^{n_{ij}^W}}{\Lambda_W}\,e^{\pm i a\,n^W_{ij}/v_\phi}\left(\overline{L_i}\tilde{H}\right)\left(\tilde{H}^T L_j^C\right)+{\rm{h.c.}}\,,
\end{align}
where $\tilde H \equiv i\sigma_2 H^\ast$, $C$ is the charge-conjugation operator, and $r_{ij}^{u,d,e}$ are $\mathcal{O}(1)$ coefficients. Here, $n_{ij}^{u,d,e}$ denote respectively $|X_{Q_i}-X_{u_j}|$, $|X_{Q_i}-X_{d_j}|$, and $|X_{L_i}-X_{e_j}|$,\footnote{The $\pm$ sign in front of the axion phase depends on whether the effective Yukawa term arises from an insertion of $\phi$ or $\phi^{\dagger}$. Crucially, this sign is correlated with the sign of the argument in $n_{ij}$, so that minus signs always eventually cancel and the rotation defined in Eq.~\eqref{eq:chiral-rotation} always removes the axion dependence from the Yukawa couplings.} while $n_{ij}^W\equiv|X_{L_i}+X_{L_j}|$. The last term in Eq.~\eqref{eq:FNL} is the LNV dimension-five Weinberg operator, which is responsible for generating active neutrino masses. The associated Wilson-coefficient matrix, $c_{\alpha\beta}$, is symmetric in flavor space, and $\Lambda_W$ parametrizes the UV scale at which the operator is generated. The Wilson coefficients are related to the PMNS matrix $U$ as follows, 
\begin{align}
    \label{eq:PMNS}
    \frac{c_{\alpha\beta}\epsilon^{n_{\alpha\beta}^W}}{\Lambda_W}=\frac{\sum_{i=1}^3 U_{\alpha i}^\ast U_{\beta i}^\ast m_i}{2v_H^2}, \qquad \expval{H}=v_H\approx 174\,\text{GeV},
\end{align}
where $m_i$ denote the active-neutrino masses. Throughout this work, it is assumed that the maximum temperature attained in the early Universe is below the mass scale of the particle(s) that generate the Weinberg operator, ensuring that the effective field theory (EFT) description remains reliably applicable. Moreover, the validity of the effective description implicitly depends on appropriate FN charge assignments for the UV fields responsible for generating the Weinberg operator. We discuss this point in detail in Appendix~\ref{appendix:FN}, where the type-I, II, and III seesaw models are analyzed. There, we show explicitly how requiring the consistency of the effective Weinberg operator description constrains the FN charges of both the SM lepton doublets and the heavy DoF introduced in the seesaw models. Moreover, we clarify how $\frac{c_{\alpha\beta}}{\Lambda_W}$ relates to the mass scale associated with the new heavy DoF, their FN charges, and couplings in the UV.

To analyze the axion-SM interactions encoded in Eq.~\eqref{eq:FNL}, it is useful to perform a local, chiral, and flavor-dependent field redefinition of the SM fermions,
\begin{align}
    \label{eq:chiral-rotation}
    Q_j&\rightarrow e^{ia(x)X_{Q_j}/v_\phi}Q_j, \quad 
    u_j\rightarrow e^{ia(x)X_{u_j}/v_\phi}u_j,\quad  
    d_j\rightarrow e^{ia(x)X_{d_j}/v_\phi}d_j,\nonumber \\  
    L_j&\rightarrow e^{ia(x)X_{L_j}/v_\phi}L_j,\quad  
    \,\,e_j\rightarrow e^{ia(x)X_{e_j}/v_\phi}e_j,
\end{align}
which rotates the axion away from the Yukawa terms and generates a derivative coupling between the axion and the FN current,
\begin{equation}
 \mathcal{L}\supset -\frac{\partial_{\mu} a(x)}{v_{\phi}}j_{\text{FN}}^{\mu}\,.
\end{equation}
Here, the FN current is given by\footnote{Note that since $X_H=0$, the axion does not couple directly to the SM Higgs.}
\begin{align}
    j^\mu_{\text{FN}}&=\sum_{i=1}^3\left[
    X_{Q_i}\Bar{Q}_{i}\gamma^\mu Q_{i}
    +X_{u_i}\Bar{u}_i\gamma^\mu u_i
    +X_{d_i}\Bar{d}_i\gamma^\mu d_i 
    +X_{L_i}\Bar{L}_{i}\gamma^\mu L_{i}
    +X_{e_i}\Bar{e}_i\gamma^\mu e_i
    \right].
\end{align}
Since the chiral rotation in Eq.~\eqref{eq:chiral-rotation} is anomalous, the field redefinitions also generate topological couplings between $a(x)$ and the gauge fields.\footnote{The coefficient is readily obtained by computing the mixed $U(1)SU(N)^2$ anomalous contribution to the divergence of the FN current.
}
The resulting axion-SM interactions are given below
\begin{align}
    \label{eq:a_lag}
    \mathcal{L}&\supset  
    -\frac{\partial_\mu a(x)}{v_\phi}j^\mu_{\text{FN}}
    -\frac{a(x)}{v_\phi}\frac{g_w^2}{32\pi^2}
    \sum_{i=1}^3 \left(X_{L_i}+3X_{Q_i}\right)
    W^a_{\mu\nu}\tilde{W}^{a\mu\nu}
    \nonumber\\
    &-\frac{a(x)}{v_\phi}\frac{g_s^2}{32\pi^2}
    \sum_{i=1}^3 \left(2X_{Q_i}-X_{u_i}-X_{d_i}\right)
    G^a_{\mu\nu}\tilde{G}^{a\mu\nu} \nonumber \\
    &\equiv-\frac{\partial_\mu a(x)}{v_\phi}j^\mu_{\text{FN}}
    -\frac{a(x)}{v_\phi}\frac{1}{32\pi^2}\left[C_w\,g_w^2W^a_{\mu\nu}\tilde{W}^{a\mu\nu}+C_s\,g_s^2G^a_{\mu\nu}\tilde{G}^{a\mu\nu}\right],
\end{align}
where we have neglected the hypercharge coupling as it does not play a direct role in the mechanism presented in this paper. It is worth noting that the FN charge assignments in Tab.~\ref{tab:wrinkles} generically predict a large value for $|C_s|$. This corresponds to a large axion-strong sphaleron coupling, as seen from Eq.~\eqref{eq:a_lag}, and makes the quark sector highly important for $B-L$ asymmetry production as discussed further in Section~\ref{sec:3}.

Beyond explaining flavor hierarchies and neutrino masses, the framework considered in this section naturally yields the BAU via spontaneous baryogenesis, as demonstrated in the rest of this paper. However, this minimal setup inherits a severe Higgs hierarchy problem and does not address the strong CP problem. In the following, we briefly outline how straightforward extensions can simultaneously resolve both issues.\\

\noindent\textbf{Higgs naturalness:} Generic couplings between the Higgs sector and heavy vector-like fermions (VLFs) with mass $M$ induce large one-loop corrections to the Higgs mass of size $\sim\frac{M^2}{16\pi^2}$. Moreover, after $U(1)_H$ is spontaneously broken, the flavon--Higgs portal operator $\phi \phi^{\dagger} H H^{\dagger}$ induces a tree-level contribution to the Higgs mass parameter of order $v_\phi^2$.\footnote{Setting the flavon--Higgs portal coupling to zero does not enhance the symmetry of the model. Consequently, even if it vanishes at a given scale, it is expected to be regenerated by radiative effects.} In the absence of additional structure, these contributions would generically destabilize the electroweak scale, thereby requiring a non-negligible degree of fine-tuning.

It is worth emphasizing, however, that the present setup admits a natural embedding within a supersymmetric framework. In such a construction, the Yukawa-like operators in Eq.~\eqref{eq:fn_lagrangian} are written in terms of the two Higgs doublets $H_u$ and $H_d$, which may carry distinct FN charges. Holomorphy of the superpotential then enforces that all exponents $n_{ij}$ appearing in Eq.~\eqref{eq:fn_lagrangian} have the same sign, implying that the effective higher-dimensional operators are generated exclusively through insertions of $\phi$, rather than $\phi^\dagger$. This corresponds to the choice $X = -Y = -1$ in Tab.~\ref{tab:wrinkles}, together with the negative sign assignment for right-handed fermions in the first generation. Supersymmetry, therefore, provides a technically natural framework in which the high-scale dynamics considered here does not destabilize the electroweak scale. The construction of explicit supersymmetric realizations is left for future work; see also the related discussion at the end of Section~\ref{sec:comparison}.\\

\noindent\textbf{Strong CP problem:} The axion considered here generally couples to $G\tilde{G}$ for generic FN charge assignments. As we will show below, explaining the BAU via standard misalignment requires a heavy axion mass (at least hundreds of TeV). Consequently, this heavy axion cannot solve the strong $CP$ problem. Dynamically explaining the smallness of $\bar{\theta}$ therefore requires an independent mechanism operating at lower scales, such as a traditional light QCD axion. An alternative approach would be to identify our axion with the QCD axion in the first place. This trade-off reduces the minimality of the baryogenesis mechanism by requiring e.g. kinetic misalignment~\cite{Co:2019jts} rather than standard misalignment and lowering the scale of LNV, cf. the discussion in Section~\ref{sec:comparison}.

\subsection{Elements of spontaneous baryogenesis}
\label{sec:SBG}
Next, we will review some elementary aspects of spontaneous baryogenesis, referring the interested reader to Refs.~\cite{Cohen:1987vi,Cohen:1988kt, DeSimone:2016ofp,Domcke:2020kcp} and references therein for additional details. Spontaneous baryogenesis holds a special place in the landscape of baryogenesis mechanisms, as it circumvents Sakharov’s condition~\cite{Sakharov:1967dj} of a departure from thermal equilibrium by invoking spontaneous breaking of $CPT$ symmetry. The latter leads to a shift in energy between particles and antiparticles, and in the presence of particle-number-violating reactions, it allows the generation of a net asymmetry between particles and antiparticles without departure from thermal equilibrium. 

The canonical setup for spontaneous baryogenesis involves an axion field $a(t)$, with decay constant $f_a$, whose background value is spatially uniform across our observable Universe. Such a homogeneous initial condition is naturally compatible with inflationary cosmology~\cite{Starobinsky:1980te,Guth:1980zm}, as long as the axion’s associated global symmetry is broken well before the end of inflation and is not restored in the post-inflationary epoch. Parametrically, this requires $f_a$ to be larger than both the Hubble scale during inflation, $H_I$, and the maximal temperature attained in the post-inflationary Universe, $T_{\text{max}}$. The second key ingredient is a derivative interaction between the axion and the current $j_Q^\mu$ of a global $U(1)_Q$ symmetry,
\begin{align}
    \label{eq:sponBG}
    \mathcal{L}_a\supset -\frac{\partial_\mu a}{f_a}j_Q^\mu=-\frac{\dot{a}}{f_a}j^0_Q=-\mu_Q^{\rm eff}j^0_Q.
\end{align}
For a time-dependent axion background, this term acts as an effective chemical potential, $\mu_Q^{\rm eff}$, for the charge associated with $U(1)_Q$. It therefore induces chemical potentials $\mu_i$ for the particle species carrying nonzero $Q$ charge.

In general, fields charged under $U(1)_Q$ may also carry charges under other global symmetries. If interactions in the thermal bath violate one such charge, denoted by $D$, the bias generated by the rolling axion background drives the plasma toward an equilibrium state with a nonzero charge density $q_D^{\rm eq}$. In the framework considered here, $U(1)_Q$ will be identified with the horizontal symmetry $U(1)_H$, while the violated charge will be taken to be $D=B-L$. The required source of $B-L$ violation is provided by the Weinberg operator, which was introduced in Eq.~\eqref{eq:FNL}. Moreover, we identify the axion-decay constant with the $U(1)_H$-breaking scale, $f_a=v_\phi$. Finally, the relation between the induced equilibrium density $q_D^{\text{eq}}$ and the effective chemical potential $\mu_Q^{\text{eff}}$ is obtained by imposing the chemical-equilibrium conditions associated with the relevant plasma processes; general expressions and explicit formulae can be found in Ref.~\cite{Domcke:2020kcp}. 

From Eq.~\eqref{eq:sponBG}, it is clear that tracking the axion velocity allows us to characterize the chemical equilibrium state of the thermal plasma, so we need to monitor its evolution. The post-inflationary displacement of the axion zero mode is parametrized as
\begin{align}
a_i = v_\phi \theta_i ,
\end{align}
where $\theta_i\in (-\pi, \,\pi]$ denotes the initial misalignment angle. In general, this angle is selected stochastically by the inflationary history. In what follows, we treat its value within our observable Hubble volume as a free parameter. After inflation, the time evolution of the spatially homogeneous axion background is determined by its classical equation of motion (EoM),\footnote{In Eq.~\eqref{EoM}, we ignore the backreaction associated with the charge densities generated in the plasma. Equivalently, we assume that any dissipative contribution from plasma interactions is much smaller than the Hubble damping term throughout the region of parameter space considered here, a point verified a posteriori.}
\begin{align}
\label{EoM}
\ddot a + 3 H(t) \dot a + \partial_a V(a) \simeq 0 .
\end{align}
For the purposes of our analysis, we expand the axion potential around one of its minima and keep only the leading quadratic term,
\begin{align}
V(a) = \frac{1}{2} m_a^2 a^2 .
\end{align}
This harmonic approximation is appropriate whenever the axion displacement relevant for the dynamics remains sufficiently far from the hilltop region, so that higher-order anharmonic terms in the potential do not play an important role. If the axion mass receives sizable thermal corrections, these effects can be included by replacing $m_a$ with a temperature-dependent quantity $m_a(T)$. In the present work, however, we restrict attention to the simpler case of a temperature-independent axion mass, with $m_a<H_I$. Possible microscopic origins of this mass term are discussed in Section~\ref{sec:cosmohistory}.

Finally, we will focus on the minimal cosmological scenario in which spontaneous LG takes place primarily during a radiation-dominated epoch after inflation. In a radiation-dominated background, the EoM in Eq.~\eqref{EoM} can be solved analytically for the homogeneous axion mode. Expressed in terms of the misalignment angle $\theta \equiv a/v_\phi$, the solution is
\begin{align}
    \theta(t)
    =
    \theta_i
    \left(\frac{2}{m_a t}\right)^{1/4}
    \Gamma\left(\frac{5}{4}\right)
    J_{1/4}(m_a t),
\end{align}
where $J_\nu$ is a Bessel function of the first kind and $\Gamma$ denotes the gamma function.
We define the characteristic temperature at which the axion begins to oscillate by the condition
\begin{align}
    \label{eq:Tosc}
    3H(T_{\rm osc}) = m_a ,
\end{align}
and introduce the dimensionless variable
\begin{align}
    z \equiv \frac{T_{\rm osc}}{T}.
\end{align}
During radiation domination, the temperature scales as $T\propto t^{-1/2}$, so that the argument of the Bessel function may be written as $m_a t \equiv c' z^2$. Using the analytic solution above, we further obtain
\begin{align}
    \label{eta(T)}
    \eta(T)
    \equiv
    \frac{\dot{\theta}(T)}{T}
    =
    -\theta_i
    \left(\frac{m_a}{T_{\rm osc}}\right)
    \left(\frac{2z^2}{c'}\right)^{1/4}
    \Gamma\left(\frac{5}{4}\right)
    J_{5/4}\left(c'z^2\right),
\end{align}
which is the quantity that enters the evolution equations discussed in Section~\ref{sec:3}.

\subsection{A viable cosmological setup}
\label{sec:cosmohistory}
\begin{figure}[th]
    \centering
    \includegraphics[width=1\textwidth]{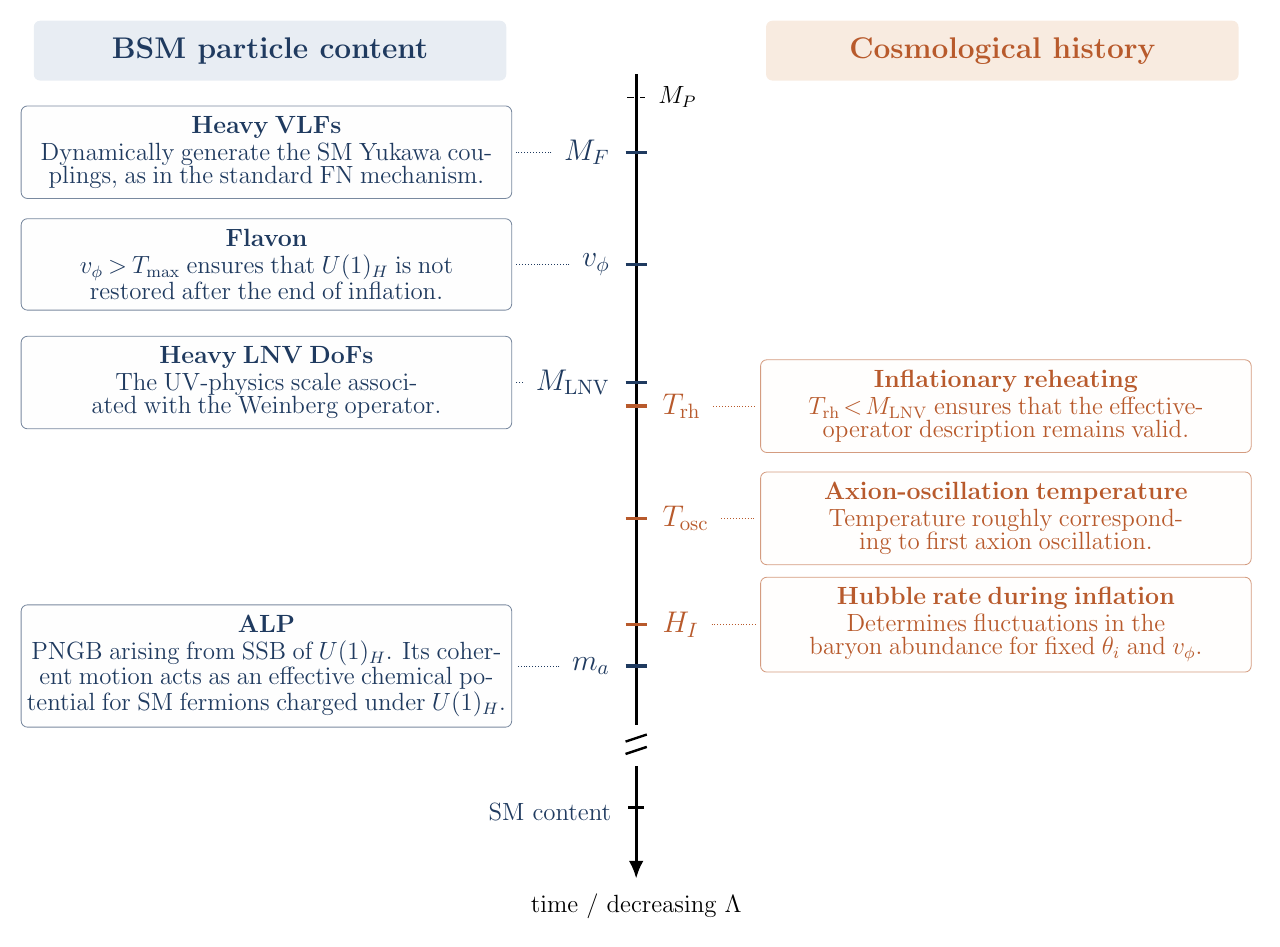}
    \caption{Schematic overview of the considered mechanism. The BSM sector consists of the standard FN model governed by a global $U(1)_H$ symmetry~\cite{Froggatt:1978nt}, augmented by unspecified degrees of freedom that generate the Weinberg operator (e.g., right-handed neutrinos in a type-I seesaw framework). The relative ordering among $M_{\text{LNV}}$ and $M_F$ as well as $M_{\text{LNV}}$ and $v_\phi$ is arbitrary; the validity of our EFT description and the absence of post-inflationary symmetry restoration simply require $M_{\text{LNV}} > T_{\text{max}}$ and $v_\phi > \max(T_{\text{max}}, H_I)$, respectively, where $T_{\text{max}}$ is the maximal temperature after the end of inflation. For simplicity, the radial mode of the $U(1)_H$-breaking scalar is assumed to have a mass comparable to $v_\phi$. This implies that $m_\phi > H_I$, so we do not expect a relevant initial displacement for the radial flavon mode, which by assumption can be integrated out from the beginning. We further assume $T_{\text{rh}} > T_{\text{osc}}$. If this condition is violated, a reliable treatment requires tracking the asymmetry generation continuously from the end of inflation, through reheating, and into the radiation-dominated epoch.}
    \label{fig:timeline}
\end{figure}

We now outline a minimal cosmological realization of the proposed framework, schematically illustrated in Fig.~\ref{fig:timeline}. We consider a post-inflationary Universe in which the only dynamical DoF relevant for the mechanism are those of the SM together with the axion, identified as the pseudo-Nambu--Goldstone boson associated with the spontaneous breaking of a global $U(1)_H$ symmetry. We assume that this symmetry is already broken during inflation and remains so throughout the subsequent thermal history. The same $U(1)_H$ symmetry is responsible for the observed flavor structure of the SM via the FN mechanism.

In addition, we include the dimension-five Weinberg operator, which generates Majorana masses for the active neutrinos. Our analysis does not depend on the specific UV completion responsible for either the Weinberg operator or the FN construction. Nevertheless, well-motivated UV realizations of the Weinberg operator are provided by the type-I, type-II, and type-III seesaw mechanisms, which we discuss in Appendix~\ref{appendix:FN}. Similarly, the FN effective operators can be generated by integrating out heavy messenger fields with masses above the $U(1)_H$-breaking scale.

Successful baryogenesis in this setup requires the axion background to be rolling while the LNV interactions induced by the Weinberg operator are still efficient. Since the latter decouple at temperatures of order $T_{W}\sim 10^{12\cdots13}\,{\rm GeV}$, the axion must acquire an appreciable velocity no later than this epoch. Equivalently, the onset of axion oscillations should occur at a temperature comparable to or above $T_{W}$. This requirement translates into a lower bound on the axion mass, as can be seen from Eq.~\eqref{eq:Tosc}.

There are several possible microscopic origins for such a mass. For instance, the axion may couple anomalously to a confining hidden gauge sector, whose nonperturbative dynamics generate a potential for the axion. A more generic possibility is that the $U(1)_H$ symmetry is explicitly broken in the UV. While the origin of this breaking may vary, a feasible scenario is that quantum-gravity effects violate the global $U(1)_H$ symmetry through Planck-suppressed operators. For instance, operators of the form
\begin{align}
    \Delta V \supset c_n \frac{\phi^n}{M_P^{n-4}} + {\rm h.c.},
    \qquad n>4 ,
\end{align}
where $\phi$ is the flavon field, generate an axion potential after $\phi$ acquires a vev. Parametrically, the resulting contribution to the axion mass is
\begin{align}
    \Delta m_a
    \sim
    n \sqrt{|c_n|}\,
    v_\phi
    \left(\frac{v_\phi}{M_P}\right)^{\frac{n}{2}-2},
\end{align}
up to order-one factors that depend on the phase of $c_n$.
If this contribution is not the dominant source of the axion mass, or if several such operators contribute with unrelated phases, avoiding cancellations or distortions of the desired potential requires each individual contribution to be at most of order the mass needed for oscillations at $T_{\rm osc}$. Imposing
$\Delta m_a \lesssim 3H(T_{\rm osc})$ gives
\begin{align}
    |c_n|
    \lesssim
    \frac{1}{n^2}
    \frac{3^2H^2(T_{\rm osc})}{v_\phi^2}
    \left(\frac{M_P}{v_\phi}\right)^{n-4}.
\end{align}
This may be written parametrically as
\begin{align}
    |c_n|
    \lesssim
    \frac{g_*(T_{\rm osc})}{n^2}
    \left(\frac{T_{\rm osc}}{v_\phi}\right)^4
    \left(\frac{M_P}{v_\phi}\right)^{n-6}.
\end{align}
Thus, unless the leading $U(1)_H$-violating operators appear only at sufficiently high dimension, their coefficients must be strongly suppressed. This is a standard axion-quality problem: Planck-suppressed operators generically break the would-be global symmetry and can either generate an axion mass that is too large 
or require tuning among unrelated contributions. Due to the large mass scales involved, the problem is much less severe than for the standard QCD axion arising from a $U(1)_{PQ}$ symmetry.

\subsection{Comparison with some related frameworks}
\label{sec:comparison}
Finally, we emphasize that the framework presented here represents only one possible avenue toward a unified understanding of the origin of flavor and the BAU. A number of alternative mechanisms have been proposed in the literature, some of which share common ingredients with our construction, while others pursue conceptually distinct approaches. In the following, we briefly review a selected subset of these scenarios, focusing on those that are closely related to our setup. We also compare our framework with alternative realizations of spontaneous baryogenesis that are similarly motivated by addressing multiple BSM puzzles within a unified setup.  \\

\noindent\textbf{Flavon carries asymmetry and decays:} Refs.~\cite{Chen:2019wnk,Elahi:2020pxl} considered a baryogenesis mechanism in which a primordial flavon--antiflavon asymmetry is converted through out-of-equilibrium flavon decays into a left--right asymmetry in the charged-lepton sector, with equal and opposite asymmetries stored in left-handed and right-handed leptons. If the flavon dominates the energy density near the electroweak epoch, the enhanced Hubble rate prevents the right-handed electrons from equilibrating before sphaleron freeze-out, allowing electroweak sphalerons to partially convert the left-handed lepton asymmetry into the BAU. In addition to requiring a specific range for the flavon mass, $\mathcal{O}(1-10)$ TeV, the mechanism requires a huge primordial flavon asymmetry of $\mathcal{O}(10^{-2}-1)$. In this sense, it is less minimal than the mechanism proposed in this paper. \\

\noindent\textbf{Flavon-assisted decay LG:} 
In the recent work of Ref.~\cite{Shao:2026xcb}, it was shown that extending the type-I seesaw framework by a flavon field may open new decay channels of the form $N_{I}\rightarrow N_J\phi$, with $I\neq J$, and introduces additional sources of CP violation~\cite{LeDall:2014too}. 
This provides a new avenue for realizing low-scale LG without relying on the mass degeneracy required in standard resonant LG~\cite{Pilaftsis:2003gt,Pilaftsis:2005rv}. 
Their approach to connecting flavor and LG is complementary to ours: Ref.~\cite{Shao:2026xcb} shows that flavon dynamics can enhance the efficiency of low-scale, decay-driven LG, whereas our mechanism can operate when the flavon is heavier than all RHNs and even in the absence of dynamical right-handed neutrinos in the early Universe.\\

\noindent\textbf{Flaxion:} In Ref.~\cite{Ema:2016ops} (see also Ref.~\cite{Calibbi:2016hwq}), the SM is extended by a horizontal symmetry, broken by a flavon, together with at least two right-handed neutrinos. 
The resulting framework addresses the strong-$CP$ problem as well as the origins of DM, the BAU, active-neutrino masses, and inflation. It is therefore considerably more ambitious in scope than the scenario considered here. 
In that model, however, the BAU is generated through conventional high-scale thermal LG, so that flavor is not directly tied to the origin of the baryon asymmetry. 
The FN breaking scale required in Ref.~\cite{Ema:2016ops} is parametrically comparable to the one in our setup, but the associated Goldstone boson is identified with the QCD axion and is at least $20$ orders of magnitude lighter than the axion considered here. Consequently, the model of Ref.~\cite{Ema:2016ops} is subject to a much more severe axion-quality problem than our scenario.\\

\noindent\textbf{Majoron-driven spontaneous LG:} 
The Majoron is the Goldstone boson associated with spontaneously broken lepton number~\cite{Chikashige:1980ui,Gelmini:1980re}. 
If it acquires a suitable mass from explicit lepton-number breaking and develops a coherent motion in the early Universe, it can drive spontaneous LG in the presence of either the Weinberg operator~\cite{Ibe:2015nfa,Domcke:2020kcp,Kuckenberg:2026oax} or the heavy dynamical DoF responsible for generating the Weinberg operator~\cite{Chun:2023eqc,Berbig:2025hlc,Chun:2025abp}. 
A major obstacle for Majoron-driven spontaneous LG, however, is the long Majoron lifetime. In standard misalignment scenarios, its late decays inject excessive entropy into the thermal bath, diluting the baryon asymmetry below the observed value~\cite{Kuckenberg:2026oax}. If kinetic misalignment is invoked~\cite{Co:2019jts}, it may even overclose the Universe~\cite{Chun:2023eqc}. 
The simplest way to avoid this problem is to introduce additional \textit{ad hoc} interactions that shorten the Majoron lifetime. 
As we will show, our scenario preserves the key advantage of Majoron-driven LG---namely, the emergence of a direct coupling between the axion velocity and the Weinberg operator---while naturally avoiding the standard Majoron-decay problem. The required additional couplings arise automatically and reduce the lifetime to an acceptable level without tuning.\\

\noindent\textbf{Axiogenesis:} Axiogenesis~\cite{Co:2019wyp} (see also Refs.~\cite{Co:2020jtv,Co:2020xlh,Barnes:2024jap}) can be viewed as a specific realization of spontaneous baryogenesis wherein a large initial axion velocity, driven by kinetic misalignment~\cite{Co:2019jts}, acts as an effective chemical potential. 
Because kinetic misalignment allows the axion to be parametrically lighter than the axion in our scenario, the axion can simultaneously serve as a DM candidate. 
The variant most closely related to our setup is lepto-axiogenesis~\cite{Co:2020jtv,Barnes:2024jap}, in which either the Weinberg operator or dynamical RHNs provide the requisite $B-L$ violation. 
While the reliance on kinetic misalignment renders this approach less minimal than ours, it offers the broader advantage of incorporating a DM explanation. 
Notably, by supersymmetrizing our framework, invoking kinetic misalignment, and introducing dynamical RHNs with appropriate masses, one could realize a model of ``lepto-axiogenesis from flavor.'' 
This would offer the intriguing theoretical benefit of rooting the lepto-axiogenesis paradigm in flavor physics, rather than relying on an \textit{ad hoc} $U(1)$ symmetry. Finally, although the axion-quality problem would be more severe in such a setup than in the scenario considered here, it would significantly improve the naturalness of the SM Higgs sector.\footnote{In fact, the flat scalar potential required in axiogenesis makes it arguably most elegantly realized with supersymmetry, which could automatically improve the naturalness of the Higgs sector. However, even in the absence of supersymmetry, Higgs naturalness is inherently improved as a direct consequence of the lowered new physics (NP) scale.}
We leave a detailed exploration of this compelling possibility for future work.

\section{Spontaneous leptogenesis from flavor}
\label{sec:3}
In this section, we first formulate the transport equations governing the evolution of particle asymmetries within our framework. We will then proceed to solve this coupled system of equations numerically, and scan over a wide range of axion-oscillation temperatures and active-neutrino masses. Among other insights, this allows us to identify the lowest possible scale for axion oscillations that remains compatible with generating a baryon asymmetry of magnitude comparable to the BAU.

The evolution of particle asymmetries generated through our mechanism can be tracked by solving a set of Boltzmann equations.  
The full network of evolution equations can be expressed in terms of the following master formula~\cite{Domcke:2020kcp},
\begin{align}
    \frac{d}{d\ln T}\left(\frac{\mu_i}{T\theta_i}\right)
    &=
    \frac{1}{g_i}
    \sum_\alpha n_i^\alpha \frac{\gamma_\alpha}{H}
    \left[
        \sum_j n_j^\alpha
        \left(\frac{\mu_j}{T\theta_i}\right)
        -
        n_S^\alpha
        \left(\frac{\eta(T)}{\theta_i}\right)
    \right] .
\end{align}
Here \(\theta_i\) denotes the initial misalignment angle. 
Normalizing the chemical potentials by \(\theta_i\) allows us to solve the system only once and subsequently obtain the baryon asymmetry for arbitrary \(\theta_i\) by a simple rescaling~\cite{Kuckenberg:2026oax}. 
An explicit expression for \(\eta(T)\) is given in Eq.~\eqref{eta(T)}, while \(g_i\) denotes the color and weak-isospin multiplicity of the corresponding SM field. 
The charge vectors \(n^\alpha\) specify how the SM fields participate in the interaction \(\alpha\); see Ref.~\cite{Domcke:2020kcp} and the Supplemental Material of Ref.~\cite{Domcke:2020quw} for details. 
The source vector \(n_S^\alpha\), in turn, encodes the coupling of the axion to the corresponding interaction. Specifically, the sphaleron couplings in Eq.~\eqref{eq:a_lag} have the following associated charge vectors
\begin{align}
    n_S^{SS} = C_s, 
    \qquad 
    n_S^{WS} = C_w ,
\end{align}
whereas the derivative coupling 
\((-\partial_\mu a/f)J_{\rm FN}^\mu\) in Eq.~\eqref{eq:a_lag} gives
\begin{align}
    n_S^\alpha = - \sum_i X_i n_i^\alpha .
\end{align}
Explicit expressions for the complete set of Boltzmann equations used in this work will be given below. 

While the general form of the Boltzmann equations can be formulated without specifying lepton-flavor indices, the physical choice of lepton basis strongly depends on the temperature regime. As is well established in the LG literature~\cite{Davidson:2008bu}, LNV processes and spectator processes sourced by charged-lepton Yukawa couplings favor distinct bases. In our case, where LNV processes are mediated by the Weinberg operator, the former aligns with the neutrino-mass basis $(1,2,3)$, whereas the latter selects the charged-lepton flavor basis $(e,\mu,\tau)$. In the temperature regime of interest to this work, only the $\tau$-Yukawa coupling is relevant; interactions mediated by the lighter lepton Yukawas remain negligible relative to the Hubble expansion rate. This reduces our viable basis choices to either $(1,2,3)$ or $(1,2,\tau)$. The $(1,2,3)$ basis provides an excellent description at $T \gg 10^{12}$ GeV (well above the $\tau$-Yukawa equilibration temperature), while the $(1,2,\tau)$ basis is appropriate for $10^{9}\text{ GeV} \ll T \ll 10^{12}$ GeV. In the intermediate transition regime at $T \sim 10^{12}$ GeV, quantum coherence effects become important. A full description of these dynamics necessitates a quantum kinetic treatment~\cite{Sigl:1993ctk,Abada:2006fw,Nardi:2006fx}, which lies beyond the scope of this paper. However, recent numerical analyses~\cite{Kuckenberg:2026oax} have demonstrated that the $(1,2,3)$ basis yields results highly consistent with the $(1,2,\tau)$ basis at $T \lesssim 10^{12}$ GeV. Motivated by this finding, we adopt the $(1,2,3)$ basis for all calculations throughout the remainder of this work. In this basis, the LNV rates mediated by the Weinberg operator read, 
\begin{align}
    \label{rate1-123}
    \gamma^{\text{eq}}_{\text{sub}}(H L_i\rightarrow H^* L_j^*)&=\frac{3T^6m_im_j}{8\pi^5v_H^4}\delta_{ij},  \\
    \label{rate2-123}
    \gamma^{\text{eq}}(L_i L_j\rightarrow H^*H^*)&=\frac{3T^6m_im_j}{16\pi^5v_H^4}\delta_{ij},
\end{align}
where $m_i$ denotes the $i$'th active neutrino mass. For concreteness, we will consider a normal-ordered neutrino hierarchy in this paper, and let the mass of the lightest active neutrino, $m_1$, vary between $0$ eV and $0.03$ eV, where the latter corresponds to the upper bound on the sum of neutrino masses $\sum_i m_i<0.12$ eV by PLANCK 2018~\cite{Planck:2018vyg}, for normal ordering. 

Following Ref.~\cite{Kuckenberg:2026oax}, we will consider the following set of chemical potentials, $\mu_i$, with\footnote{Here, we are discarding $\mu_\tau$, to keep a consistent description based on the $(1,2,3)$ basis~\cite{Kuckenberg:2026oax}. Moreover, $\mu_{q_{uds}}$ is the common chemical potential of the right-handed $u,\, d,\, s$ (with multiplicity factor $g_i=9$).} 
\begin{align}
\label{eq:i=}
i=L_1,\,L_2,\,L_3,\,t,\,c,\,b,\,q_{uds},\,Q_1,\,Q_2,\,Q_3,\,H.
\end{align}
This is a sufficient description at temperatures $T\gtrsim 10^{10}$ GeV, where SM interactions are too slow to resolve the three lightest right-handed quarks and the two lightest right-handed leptons. However, in our setup a further complication arises because the charges of Tab.~\ref{tab:wrinkles} are assigned to gauge eigenstates $L_G$, i.e. to the basis in which $U(1)_H$ acts diagonally. The fields $L_i$ entering the Boltzmann equations are instead
neutrino mass eigenstates, in which the LNV rates~\eqref{rate1-123}-\eqref{rate2-123} are diagonal. Mass and gauge eigenstates are related as $L_G=\sum_iV_{Gi}L_i$, with $V$ the unitary matrix that diagonalizes the Weinberg operator in the FN basis, so the axion current $\sum_G X_{L_G}\bar L_G\gamma^\mu L_G$ is not diagonal in the mass basis: The charge matrix there is $\tilde X_L=V^\dagger\,\mathrm{diag}(X_{L_G})\,V$, whose diagonal entries $[\tilde X_L]_{ii}=\sum_G|V_{Gi}|^2X_{L_G}$ are the weighted averages that source $\mu_{L_i}$ in the Boltzmann equations below. The overlaps $|V_{Gi}|^2$ are not directly measurable: the PMNS matrix relates mass eigenstates to the charged-lepton basis, not to the FN basis. The two bases coincide up to $\mathcal O(\epsilon)$, however. Since the charged-lepton spectrum is
FN-hierarchical, the left rotation diagonalizing $Y^e_{GG'}\sim\epsilon^{|X_{L_G}-X_{e_{G'}}|}$ has angles $\epsilon^{|X_{L_G}-X_{L_{G'}}|}$, cf.\ Eq.~\eqref{eq:VCKM}: it mixes the first generation with the other two only at
$\mathcal O(\epsilon^{|Y|})=\mathcal O(\epsilon)$, while its $\mathcal O(1)$ rotation in the $2$--$3$ plane leaves $X_L$ invariant because the second and third generations carry the same charge, $l_0$. We therefore set $V=U$, identifying the FN basis with the charged-lepton basis, which yields $\tilde X_L=U^\dagger\,\mathrm{diag}(X_{L_e},X_{L_\mu},X_{L_\tau})\,U$ with $X_{L_\alpha}$ the charges of Tab.~\ref{tab:wrinkles}, up to an $\mathcal O(\epsilon)$ uncertainty from the neglected charged-lepton rotation. The off-diagonal entries of $\tilde X_L$ source flavor coherences that diagonal chemical potentials cannot describe and are dropped, consistently with the $(1,2,3)$-basis treatment,\footnote{These off-diagonal entries are sizeable for both benchmarks considered below, so dropping them is an approximation, to be validated by the quantum kinetic treatment mentioned above.} and we define
\begin{align}
    \label{eq:TildeX}
    \tilde X_{L_i}\equiv\big[U^\dagger\,\mathrm{diag}(X_{L_e},X_{L_\mu},X_{L_\tau})\,U\big]_{ii}
    =\sum_{\alpha=e,\mu,\tau}|U_{\alpha i}|^2\,X_{L_\alpha}\,.
\end{align}
The same rotation applies to the quark-sector Yukawa sources entering the Boltzmann equations below, which are written in the quark mass basis, but there it is negligible. With hierarchical masses, the left and right rotations diagonalizing the quark Yukawa matrices have angles $\theta_{ij}\sim\epsilon^{|X_i-X_j|}$, the same counting that yields the CKM matrix in
Eq.~\eqref{eq:VCKM}. The fields entering the sources, $Q_{2,3}$, $c$, $t$ and $b$, differ in charge from every other quark of the same type by at least one unit, so all relevant angles are at most $\epsilon$ and the diagonal charges shift only by terms of order $\epsilon^2\simeq0.04$. In particular, the top-Yukawa source, which vanishes exactly in the FN basis, becomes $\mathcal O(\epsilon^2)$. This is far below the basis-independent sphaleron sources $C_s$ and $C_w$ that dominate the quark sector.

The axion does not only couple to Yukawa interactions whose couplings are generated dynamically by the flavon vev;
it also generically couples to weak sphalerons, strong sphalerons, and the LNV processes mediated by the Weinberg operator, as can be seen clearly from the Boltzmann equations presented in Section~\ref{sec:BEqs}. 
To disentangle the roles of these different processes, we will therefore organize the analysis of the evolution equations in three steps. 
We first consider the case in which only the quark-sector FN charges are nonzero, and then study the opposite case in which only the lepton-sector FN charges are nonzero. In the end, we perform a full analysis with nonzero FN charges in both sectors. 
For concreteness, we focus on two representative benchmark choices of FN charges, listed in Tab.~\ref{tab:charges}, that reproduce the observed masses and flavor textures of the SM quark and lepton sectors and are compatible with a consistent EFT description of the Weinberg operator, cf. Appendix~\ref{appendix:FN}. Meanwhile, the axion oscillation temperature is treated as a free parameter. 

\begin{table}[htbp]
\centering
\rowcolors{2}{white}{Orange!8}
\begin{tabular}{c ccc ccc ccc ccc ccc cc}
\toprule
Benchmark & $Q_1$ & $Q_2$ & $Q_3$ & $ u_1$ & $ u_2$ & $ u_3$ & $ d_1$ & $ d_2$ & $ d_3$ & $ L_1$ & $ L_2$ & $ L_3$ & $ e_1$ & $ e_2$ & $ e_3$ & $C_w$ & $C_s$  \\
\midrule
BP1  &3  &2  &0  &-4  &-1  &0  &-3 &-3 &-2 &1 &0 &0 &-7 &-5 &-3 & 16 &23  \\
BP2  &4  &3  &1  &-3  &0  &1  &-2 &-2 &-1 &0 &-1 &-1 &-8 &-6 &-4 & 22 &23   \\
\bottomrule
\end{tabular}
\caption{FN charges and the resulting anomaly coefficients for the two benchmark scenarios considered below. In the general
parametrization of Tab.~\ref{tab:wrinkles}, BP1 corresponds to $X=-1,\, Y=1,\, l_0=0,\, q_0=0,$ while BP2 corresponds to $X=-1,\, Y=1,\, l_0=-1,\, q_0=-1.$ The charges are assigned to gauge eigenstates, which in general differ from the eigenstates entering our Boltzmann equations, as discussed in the text below Eq.~\eqref{eq:i=}.}
\label{tab:charges}
\end{table}

\subsection{Boltzmann equations}
\label{sec:BEqs}

We now present explicit expressions for the full set of Boltzmann equations for the most general case of high-scale spontaneous LG from $U(1)_H$, with arbitrary FN charge assignments. 
The chemical potential associated with the SM Higgs is affected by all Yukawa interactions and LNV interactions induced by the Weinberg operator,  
\begin{align}
\label{eq:muH_full}
\frac{d}{d\ln T}\left(\frac{\mu_H}{T}\right) &=
\frac{\gamma_t(T)}{4H(T)}
\left[\left(\frac{-\mu_t + \mu_{Q_3} + \mu_H}{T} \right)+\eta(T)\left(-X_t + X_{Q_3} + X_H\right)\right]\nonumber\\
&- \frac{\gamma_b(T)}{4H(T)}\left[\left(\frac{-\mu_b + \mu_{Q_3} - \mu_H}{T} \right)+\eta(T)\left(-X_b + X_{Q_3} - X_H\right)\right]\nonumber \\
&+ \frac{\gamma_c(T)}{4H(T)}
\left[\left(\frac{-\mu_c + \mu_{Q_2} + \mu_H}{T} \right)+\eta(T)\left(-X_c + X_{Q_2} + X_H\right)\right]
\nonumber \\
&+\frac{1}{2}\sum_{i}\left(\frac{\gamma^{\text{eq}}_{\text{sub}}(HL_i\rightarrow H^*L_i^*)+2\gamma^{\text{eq}}(L_i L_i\rightarrow H^*H^*)}{{(T^3/6)H}}\right)\nonumber \\
    &\times \left(\frac{2\mu_H+2\mu_{L_i}}{T}+2(\tilde{X}_{L_i}+X_H)\eta(T)\right).
\end{align}
The quark chemical potentials are directly affected by Yukawa interactions, and both strong and weak sphalerons,
\begin{align}
\frac{d}{d\ln T}\left(\frac{\mu_t}{T}\right) &=-\frac{\gamma_{\text{SS}}(T)}{3H(T)}\left(\frac{-3\mu_{q_{uds}} - \mu_t-\mu_c - \mu_b+ 2\mu_{Q_1}+2\mu_{Q_2} + 2\mu_{Q_3}}{T}
- C_s\,\eta(T)\right)\nonumber\\
&
-\frac{\gamma_t(T)}{3H(T)}\left[\left(\frac{-\mu_t + \mu_{Q_3}+ \mu_H}{T}\right)+\eta(T)\left(-X_t + X_{Q_3}+ X_H\right)\right],
\\[0.8cm]
\frac{d}{d\ln T}\left(\frac{\mu_c}{T}\right) &=-\frac{\gamma_{\text{SS}}(T)}{3H(T)}\left(\frac{-3\mu_{q_{uds}} - \mu_t - \mu_b - \mu_c+ 2\mu_{Q_{1}} +2\mu_{Q_{2}}+ 2\mu_{Q_3}}{T}
- C_s\,\eta(T)\right)\nonumber\\
&
-\frac{\gamma_c(T)}{3H(T)}\left[\left(\frac{-\mu_c + \mu_{Q_2} + \mu_H}{T} \right)+\eta(T)\left(-X_c + X_{Q_2} + X_H\right)\right],
\\[0.8cm]
\frac{d}{d\ln T}\left(\frac{\mu_b}{T}\right) &=-\frac{\gamma_{\text{SS}}(T)}{3H(T)}\left(\frac{-3\mu_{q_{uds}} - \mu_t-\mu_c - \mu_b+ 2\mu_{Q_1}+2\mu_{Q_2} + 2\mu_{Q_3}}{T}
- C_s\,\eta(T)\right)\nonumber\\
&-\frac{\gamma_b(T)}{3H(T)}\left[\left(\frac{-\mu_b + \mu_{Q_3} - \mu_H}{T} \right)+\eta(T)\left(-X_b + X_{Q_3} - X_H\right)\right],\\[0.8cm]
\frac{d}{d\ln T}\left(\frac{\mu_{q_{uds}}}{T}\right) &=-\frac{\gamma_{\text{SS}}(T)}{3H(T)}
\left(\frac{-3\mu_{q_{uds}} - \mu_t-\mu_c - \mu_b+ 2\mu_{Q_1}+2\mu_{Q_2} + 2\mu_{Q_3}}{T}- C_s\,\eta(T)\right),\\[0.8cm]
\frac{d}{d\ln T}\left(\frac{\mu_{Q_i}}{T}\right) &=\frac{\gamma_{\text{WS}}(T)}{2H(T)}\left(\frac{\mu_{L_1} +\mu_{L_2}+ \mu_{L_3} + 3\mu_{Q_{1}} +3\mu_{Q_{2}} + 3\mu_{Q_3}}{T}- C_w\,\eta(T)\right)\nonumber\\&
+ \frac{\gamma_{\text{SS}}(T)}{3H(T)}
\left(\frac{-3\mu_{q_{uds}} - \mu_t-\mu_c - \mu_b+ 2\mu_{Q_1}+2\mu_{Q_2} + 2\mu_{Q_3}}{T}- C_s\,\eta(T)\right)\nonumber \\
& + \delta_{i2}\frac{\gamma_c(T)}{6H(T)}\left[
\left(\frac{-\mu_c + \mu_{Q_2} + \mu_H}{T} \right)+\eta(T)\left(-X_c + X_{Q_2} + X_H\right)\right]\nonumber \\
&+\delta_{i3} \frac{\gamma_t(T)}{6H(T)}
\left[\left(\frac{-\mu_t + \mu_{Q_3} + \mu_H}{T} \right)+\eta(T)\left(-X_t + X_{Q_3} + X_H\right)\right]\nonumber \\
&+\delta_{i3} \frac{\gamma_b(T)}{6H(T)}\left[\left(\frac{-\mu_b + \mu_{Q_3} - \mu_H }{T}\right)+\eta(T)\left(-X_b + X_{Q_3} - X_H \right)\right].
\end{align}
The lepton chemical potentials are directly affected by electroweak sphalerons and LNV interactions induced by the Weinberg operator,\footnote{Recall that, working in the $(1,2,3)$ basis, we ignore effects induced by the $\tau$-Yukawa coupling.}  
\begin{align}
\label{eq:muL_full}
\frac{d}{d\ln T}\left(\frac{\mu_{L_i}}{T}\right)&=\frac{\gamma_{WS}(T)}{2H(T)}
\left(\frac{\mu_{L_{1}}+\mu_{L_{2}} + \mu_{L_{3}}
+3\mu_{Q_{1}} + 3\mu_{Q_{2}} + 3\mu_{Q_{3}}}{T}- C_w\,\eta(T)\right)
\nonumber\\
&+\left(\frac{\gamma^{\text{eq}}_{\text{sub}}(HL_i\rightarrow H^*L_i^*)+2\gamma^{\text{eq}}(L_i L_i\rightarrow H^*H^*)}{{(T^3/6)H}}\right)\nonumber \\
    &\times \left(\frac{2\mu_H+2\mu_{L_i}}{T}+2(\tilde{X}_{L_i}+X_H)\eta(T)\right).
\end{align} 
Explicit expressions for all interaction rates entering these equations, as well as the values of the various couplings employed in our numerical analysis, can be found in Ref.~\cite{Kuckenberg:2026oax}. Finally, we note that the equations above rely on the linear approximation, $\mu_i/T\ll 1$, a condition we verify holds throughout our analysis.\footnote{This condition also ensures that thermal effects dominate over finite-density effects, such that thermal restoration of the electroweak vacuum dominates over (potential) density-induced breaking effects~\cite{Benson:1991nj,Hook:2019vcn}.}

\subsection{Asymmetry generation}
\label{sec:asymgen}
In the rest of this paper, we analyze the representative benchmark scenarios of Tab.~\ref{tab:charges}. These FN charges allow for an analysis that is consistent with both LNV UV physics respecting $U(1)_H$ and a valid effective description of LNV in terms of the Weinberg operator, as explained in Appendix~\ref{appendix:FN}. The doublet charges entering the Boltzmann equations follow from Eq.~\eqref{eq:TildeX}. With the PMNS parameters of Ref.~\cite{Esteban:2024eli}, they read
\begin{align}
    \label{eq:XLtilde}
    \tilde{X}_{L_i}=(0.68,\, 0.30,\, 0.02), \qquad \tilde{X}_{L_i}=(-0.32,\, -0.70,\, -0.98),
\end{align}
for BP1 and BP2, respectively.

For fixed values of $T_{\text{osc}}$ and $m_1$, the evolution equations above can be solved numerically for a given BP. Our numerical analysis assumes that the inflationary reheating temperature is significantly above the axion-oscillation temperature, $T_{\text{rh}}\gg T_{\text{osc}}$. If this hierarchy of scales is not maintained, a model-dependent treatment is instead required for a reliable result, in which asymmetry generation is tracked from the end of inflation through the reheating epoch and the subsequent onset of radiation domination. Such an analysis lies beyond the scope of this work. \\

\begin{figure}[h]%
    \centering
    {{\includegraphics[width=0.494\textwidth]{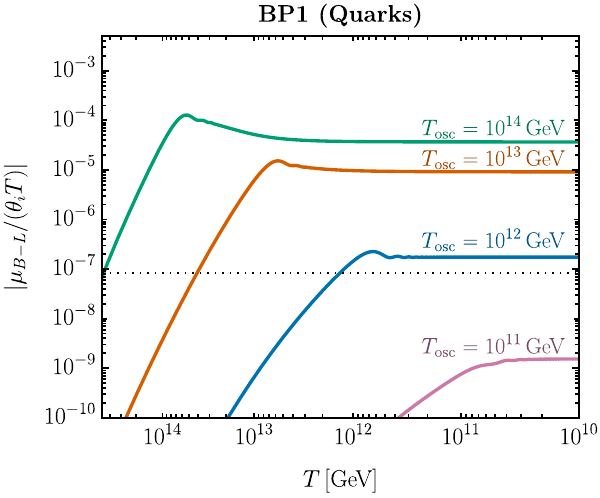}}}\, {{\includegraphics[width=0.494\textwidth]{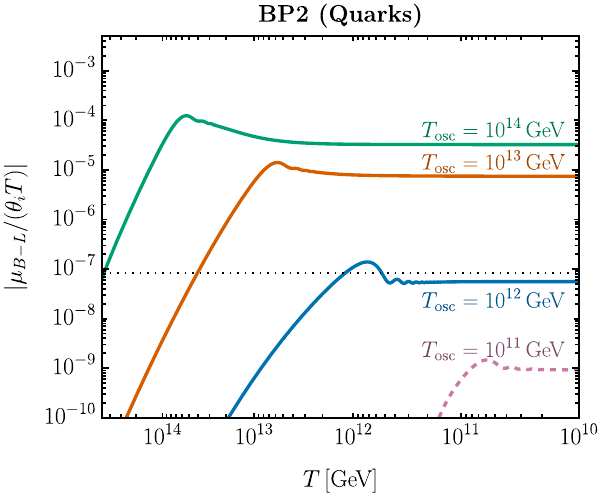}}}
    
    \caption{Evolution of $\abs{\mu_{B-L}(T)/(\theta_i T)}$ as a function of temperature for BP1 (left panel) and BP2 (right panel) with nonzero FN charges for quarks only (and FN charges for all leptons set to zero). Green, orange, blue, and purple curves correspond to $T_{\text{osc}}=10^{14},$ $10^{13},$ $10^{12},$ $10^{11}$ GeV, respectively. Solid/ dashed curves indicate where $\mu_{B-L}(T)/(\abs{\theta_i} T)$ is positive/ negative. The horizontal dotted line indicates the $B-L$ asymmetry corresponding to the BAU, normalized by $\theta_i$.}
    \label{fig:BPEvolQuarks}
\end{figure}

\noindent\textbf{Quark analysis:} We first consider the benchmark scenarios outlined in Tab.~\ref{tab:charges}, but with all lepton FN charges set to zero. Additionally, we fix the lightest neutrino mass to $m_1 = 0.01$~eV. Solving the system of Boltzmann equations for $T_{\text{osc}} = 10^{14}, 10^{13}, 10^{12},$ and $10^{11}$~GeV yields the results displayed in Fig.~\ref{fig:BPEvolQuarks}, where the left (right) panel corresponds to BP1 (BP2). The green, orange, blue, and purple curves show our results for the different oscillation temperatures, with solid (dashed) lines indicating a positive (negative) value of $\mu_{B-L}/(\abs{\theta_i}T)$. The horizontal black dotted curve indicates
\begin{align}
    \label{eq:analytic2}
    \frac{\mu_{B-L}^{\text{obs}}(T)}{\abs{\theta_i}T}=c_{\text{sph}}^{-1}\,\frac{4\pi^2g_{*s}}{15\abs{\theta_i}}\,Y_B^{\text{obs}} \,,
\end{align}
where $c_{\text{sph}}$ is the sphaleron-conversion factor~\cite{Harvey:1990qw,Laine:1999wv} and $Y_B^{\text{obs}}\simeq 8.7\cdot 10^{-11}$~\cite{Planck:2018vyg} is the yield corresponding to the observed BAU. 

For the highest oscillation temperatures (green and orange curves), only the top-quark Yukawa and strong-sphaleron interactions are fully or nearly equilibrated for $T \gtrsim T_{\rm osc}$. Since $-X_t + X_{Q_3} + X_H = 0$ for
both benchmarks, the source term associated with top-Yukawa interactions vanishes identically, and asymmetry generation in this regime is driven almost entirely by strong sphalerons. Weak-sphaleron interactions and the remaining Yukawa interactions do carry non-vanishing source terms, but their rates are well below the Hubble rate at these temperatures, and they therefore play no significant role. Because $C_s$ takes the same value for both benchmarks, the resulting asymmetry production is similar for $T_{\rm osc} \gtrsim 10^{13}$~GeV.

At lower oscillation temperatures, weak sphalerons and the lighter-quark Yukawa interactions come into equilibrium and begin to affect the evolution of chemical potentials. Weak sphalerons are the decisive ingredient here: since $C_w$ differs between BP1 and BP2, the two benchmarks are driven towards different quasi-equilibrium configurations of the chemical potentials, and hence towards different final asymmetries. This effect is important for $T_{\rm osc} \lesssim T^{\rm WS}_{\rm eq} \simeq 2.5\times10^{12}$~GeV, with $T^{\rm WS}_{\rm eq}$ the weak-sphaleron equilibration temperature~\cite{Moore:2010jd,Domcke:2020kcp}, and can have a substantial impact on the $B-L$ asymmetry, as seen by comparing the blue and purple curves across the two panels.\\

\begin{figure}[h]%
    \centering
    {{\includegraphics[width=0.494\textwidth]{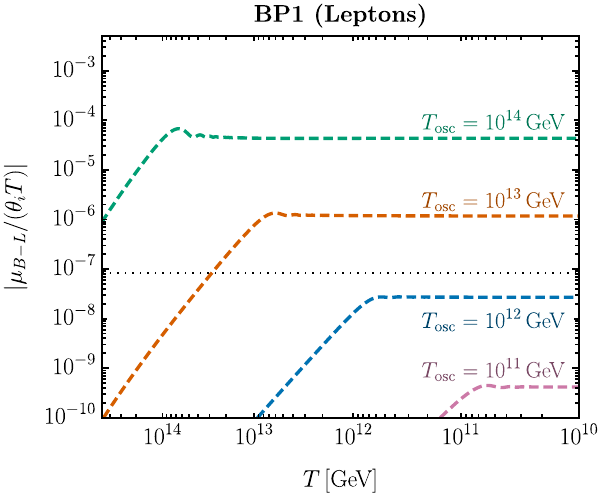}}}\, {{\includegraphics[width=0.494\textwidth]{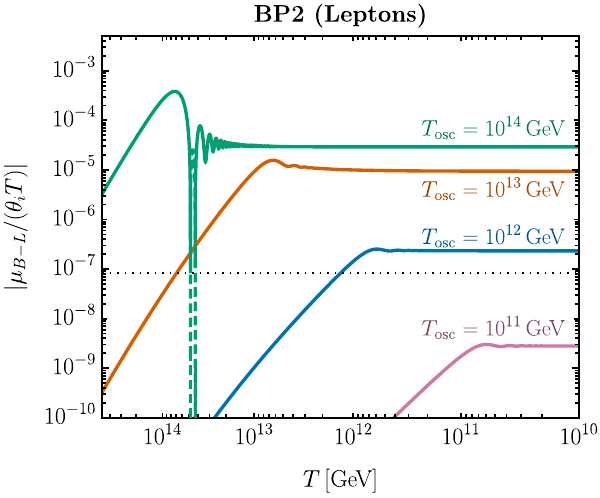}}}
    
    \caption{Similar to Fig.~\ref{fig:BPEvolQuarks}, but this time with nonzero FN charges for leptons only (and FN charges for all quarks set to zero).}
    \label{fig:BPEvolLeptons}
\end{figure}

\noindent\textbf{Lepton analysis:} We now repeat the preceding analysis, but set the FN charges of all quarks to zero while assigning FN charges to the leptons as in Tab.~\ref{tab:charges}, with the corresponding values for $\tilde{X}_L$ given in Eq.~\eqref{eq:XLtilde}. In this case, BP1 and BP2 yield qualitatively distinct results for all values of $T_{\text{osc}}$, for several reasons. First, the nonzero FN charges associated with $L_i$ possess opposite signs in the two scenarios, which is the reason why the generated asymmetries have opposite signs. Second, for a lightest neutrino mass of $m_1 = 0.01$~eV, the LNV rate 
\begin{align}
    \gamma_{\alpha\beta}^{\text{LNV}} =\gamma_{\beta\alpha}^{\text{LNV}} = \gamma^{\text{eq}}_{\text{sub}}(H L_\alpha \to H^* L_\beta^*) + (1 + \delta_{\alpha\beta}) \gamma^{\text{eq}}(L_\alpha L_\beta \to H^* H^*) \,,
\end{align}
which enters the Boltzmann equations for $\mu_H$ and $\mu_{L_i}$, satisfies $\gamma_{33}^{\text{LNV}} \gg \gamma_{22}^{\text{LNV}} >  \gamma_{11}^{\text{LNV}}$. As a result, the chemical potentials track the axion motion much more closely in BP2 than in BP1. In particular, for $T_{\text{osc}}=10^{13\cdots14}$ GeV, efficient LNV and top-Yukawa interactions force $\mu_{L_{2,3}}$, $\mu_H$, $\mu_{Q_3}$, and $\mu_t$ to closely follow the initial axion oscillations in the BP2 scenario. This distinct tracking behavior is clearly reflected in the evolution of the $B-L$ asymmetry for $T_{\text{osc}} = 10^{14}$~GeV, shown in the right panel of Fig.~\ref{fig:BPEvolLeptons}. 

For lower oscillation temperatures—depicted by the orange, blue, and purple curves—the net asymmetry generated in BP2 is significantly larger than the corresponding results for BP1. This behavior stems from the fact that the FN charge assignment in BP2 induces a stronger LNV source term; this forces the chemical potentials to track the axion velocity more tightly than in BP1, thereby maximizing asymmetry production.\\

\begin{figure}[h]%
    \centering
    {{\includegraphics[width=0.494\textwidth]{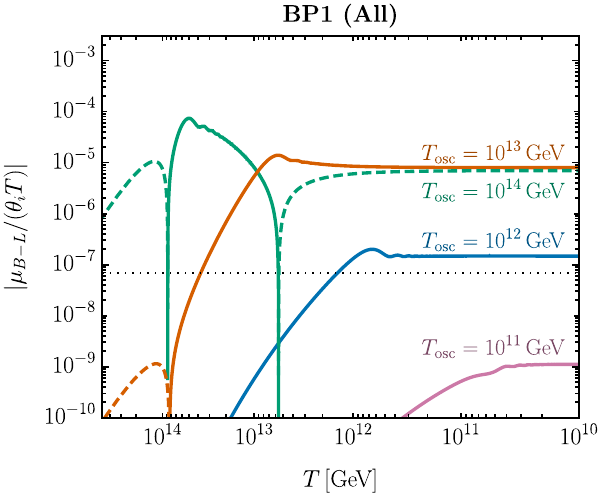}}}\, {{\includegraphics[width=0.494\textwidth]{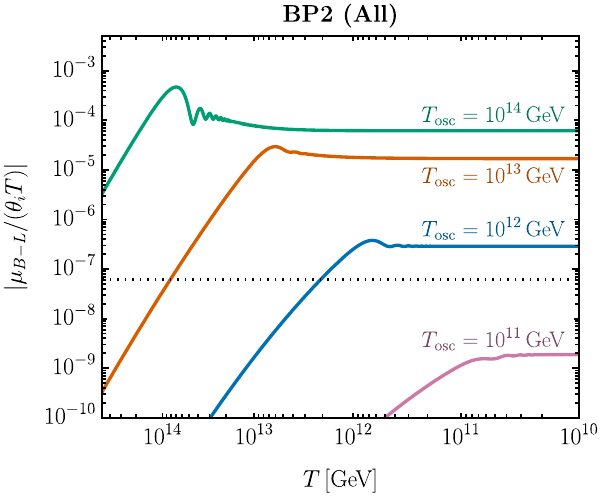}}}
    \caption{Similar to Figs.~\ref{fig:BPEvolQuarks} and~\ref{fig:BPEvolLeptons}, but this time all FN charges take the values assigned in Tab.~\ref{tab:charges}.}
    \label{fig:BPEvolAll}
\end{figure}

\noindent\textbf{Combined analysis:} 
Let us now consider the full scenario in which all SM fermions carry the FN charges assigned in Tab.~\ref{tab:charges}. We find that the initial asymmetry generation at $T \gtrsim 10^{14}$~GeV is driven primarily by the lepton doublets,
whose chemical potentials are sourced by the axion through LNV interactions, before strong sphalerons take over as the dominant effect. This is particularly clear for the green and orange curves in the left panel of Fig.~\ref{fig:BPEvolAll}, which show an asymmetry that is initially negative due to leptonic effects and subsequently changes sign once strong sphalerons come to dominate.

The interplay between the two sectors depends on both the benchmark and the oscillation temperature. For BP1 with $T_{\rm osc} = 10^{14}$~GeV, the quark and lepton contributions remain comparable in magnitude throughout the evolution and
partially cancel, so that the asymptotic $B-L$ asymmetry is suppressed by nearly one order of magnitude relative to the quark-only and lepton-only results of Figs.~\ref{fig:BPEvolQuarks} and~\ref{fig:BPEvolLeptons}. For BP2 at the same oscillation temperature the two sectors instead act constructively, and the combined asymmetry exceeds either result obtained in isolation.

At lower oscillation temperatures the picture simplifies. For BP1 with $T_{\rm osc} \lesssim 10^{13}$~GeV, the leptonic contribution is subdominant and negative, so the full result is only slightly suppressed relative to the case of
vanishing lepton charges. For BP2, by contrast, leptons provide the leading contribution to the asymptotic $B-L$ asymmetry, consistent with the stronger LNV source term discussed above.

\begin{figure}[h]%
    \centering
    {{\includegraphics[width=0.494\textwidth]{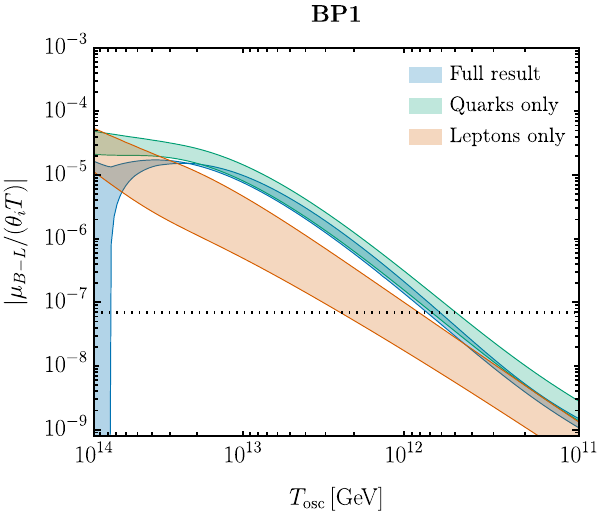}}}\, {{\includegraphics[width=0.494\textwidth]{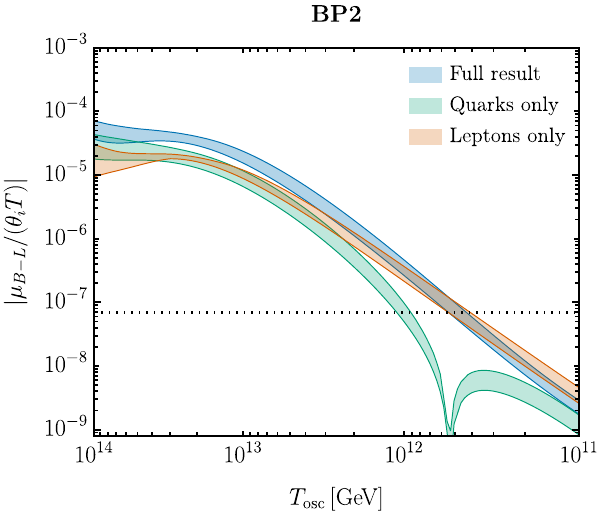}}}
    \caption{Bands showing the range within which the generated asymmetry varies as $m_1\in [0,\,0.03]$ eV is varied, consistent with the upper bound on the sum of neutrino masses from PLANCK~\cite{Planck:2018vyg} assuming a normal-ordered neutrino spectrum. The left (right) panel displays the results for BP1 (BP2). The blue, green, and orange bands correspond to the full FN charge configuration of Tab.~\ref{tab:charges}, a quark-only assignment (with vanishing lepton charges), and a lepton-only assignment (with vanishing quark charges), respectively.}
    \label{fig:ParameterBefore}
\end{figure}

Finally, Fig.~\ref{fig:ParameterBefore} provides a comprehensive overview of $B-L$ asymmetry generation as a function of oscillation temperature for BP1 (left panel) and BP2 (right panel). Here, blue, green, and orange bands correspond to the FN charges in Tab.~\ref{tab:charges}, a quark-only assignment (with vanishing lepton charges), and a lepton-only assignment (with vanishing quark charges), respectively. The bands are obtained by varying $m_1\in [0,\,0.03]$ eV. A few comments are in order: 
\begin{itemize}
    \item The figure demonstrates that our mechanism is capable of generating an asymmetry that is similar to or larger than the BAU for $T_{\text{osc}}\gtrsim \text{few}\cdot 10^{11}$ GeV. 
    \item The orange bands are similar in size to, or wider than the green bands due to how the quark and lepton sectors couple to the axion background. For left-handed leptons, LNV interactions provide a direct driving force for their chemical potentials, sourced by the axion velocity. These potentials are thereby driven directly by the LNV rates, whose magnitudes at any given temperature are controlled by $m_1$. This induces a strong sensitivity of the leptonic chemical potentials to the precise value of $m_1$. The effect of varying $m_1$ is particularly dramatic for BP1, where $\tilde{X}_{L_1}=0.68>\Tilde{X}_{L_2}=0.3\gg \Tilde{X}_{L_3}=0.02$, and the LNV rate coupling to $\mu_{L_1}$ in the Boltzmann equations is proportional to $m_1^2$; $\gamma_{11}^{\text{LNV}}\propto m_1^2$. For BP2, where $\abs{\tilde{X}_{L_1}}=0.32<\abs{\Tilde{X}_{L_2}}=0.7< \abs{\Tilde{X}_{L_3}}=0.98$, the effect is less dramatic since the LNV rates coupling to $\mu_{L_2}$ and $\mu_{L_3}$ in the Boltzmann equations do not vanish in the limit $m_1\rightarrow 0$. Moreover, the value of $\gamma_{33}^{\text{LNV}}$ is significantly less sensitive to the value of $m_1$ compared to $\gamma_{22}^{\text{LNV}}$ and in particular $\gamma_{11}^{\text{LNV}}$. Meanwhile, since LNV rates do not directly drive quark chemical potentials, the magnitude of these rates becomes a secondary effect when nonzero FN charges are assigned to quarks only, generally leading to narrower bands. 
   \item For BP1 and $T_{\rm osc}\gtrsim 8\times10^{13}$~GeV, the leptonic and quark contributions to the $B-L$ asymmetry enter with opposite signs and can be comparable in magnitude; this competition, and the resulting cancellation, were already visible for $m_1=0.01$~eV in Figs.~\ref{fig:BPEvolQuarks}--\ref{fig:BPEvolAll}. In this regime, varying $m_1$ within $[0,\,0.03]$~eV induces a sign change in the net asymptotic asymmetry, which therefore vanishes at an intermediate value. This is the origin of the large blue region on the left of Fig.~\ref{fig:ParameterBefore}.
\end{itemize}
In summary, our findings quantify the asymmetry generated through spontaneous LG for the BPs in Tab.~\ref{tab:charges}, highlighting that our framework can readily (over)produce the required baryon asymmetry over a broad range of axion-oscillation scales.

\section{Cosmological constraints}
\label{sec:4}
In the previous section, we computed the baryon asymmetry generated by our mechanism for two benchmark scenarios. Notably, the resulting asymmetry depends only on the initial misalignment angle $\theta_i = a_i/v_\phi$, rather than directly on the decay constant $v_\phi$. We now incorporate additional constraints that arise when embedding spontaneous LG into a standard cosmological history. Specifically, we must account for entropy dilution from late-time axion decays and limits from isocurvature perturbations, both of which are highly sensitive to the precise value of $v_\phi$~\cite{Kusenko:2014uta,Kuckenberg:2026oax}. Below, we summarize how to include these effects and identify the regions of the $T_{\rm osc}$--$v_\phi$ parameter space that successfully reproduce the observed BAU. \\

\noindent\textbf{Entropy dilution:} The final \(B-L\) asymmetry remaining after entropy dilution induced by axion decays can be written as~\cite{Kusenko:2014uta,Kuckenberg:2026oax}
\begin{align}
    Y_{B-L} = \Delta_a\, Y_{B-L}^{\rm sLG} ,
\end{align}
where \(\Delta_a \leq 1\) denotes the entropy-dilution factor and \(Y_{B-L}^{\rm sLG}\) is the asymmetry generated by spontaneous leptogenesis before dilution, as calculated in Section~\ref{sec:3}. 
If the axion never comes to dominate the energy density of the Universe, the associated entropy injection is negligible, and one has \(\Delta_a \simeq 1\).

In the limit \(m_a \gg v_H\), which is the regime relevant for our analysis, the axion decay rate in our setup is typically dominated by decays into gauge bosons. 
The total decay rate is then approximately
\begin{align}
    \Gamma_a 
    \simeq 
    \Gamma_a^{WW}+\Gamma_a^{gg}
    =
    \left(\frac{C_w g_w^2}{32\pi^2}\right)^2
    \frac{3}{4\pi}\frac{m_a^3}{v_\phi^2}
    +
    \left(\frac{C_s g_s^2}{32\pi^2}\right)^2
    \frac{2}{\pi}\frac{m_a^3}{v_\phi^2} .
\end{align}
Following Ref.~\cite{Kuckenberg:2026oax}, this gives the entropy-dilution factor\footnote{This result is obtained by using a decay approximation where the axion abundance decays instantaneously when the age of the Universe equals the axion lifetime. A more refined treatment would require solving coupled Boltzmann equations for the energy density stored in axions and the energy density stored in radiation.}
\begin{align}
    \Delta_a \simeq 
    \min\Bigg\{
    1,\,
    &\sqrt{\left[
        \frac{2.6\times 10^{-3}\, C_w}{\theta_i^2}
        \left(\frac{g_w}{0.55}\right)^2\right]^2
        +
        \left[\frac{5.1\times 10^{-3}\, C_s}{\theta_i^2}
        \left(\frac{g_s}{0.6}\right)^2
    \right]^2}
    \nonumber \\
    &\times \left(\frac{10^{15}\,{\rm GeV}}{v_\phi}\right)^3
    \left(\frac{m_a}{10^9\,{\rm GeV}}\right)\left(\frac{g_*(T_{\text{osc}})}{g_*(T_{{\text{dec}}})}\right)^{1/4}
    \Bigg\} ,
\end{align}
where the gauge couplings are evaluated at\footnote{In our numerical analysis, we evolve couplings using their one-loop renormalization group equations (RGEs) starting from their SM values at the $Z$ pole and run them up to the RGE scale $\mu=2\pi T$.}
\begin{align}
    \label{eq:Tdec}
    T_{\text{dec}}=\sqrt{\frac{2M_P\Gamma_a}{3}}\left(\frac{90}{g_*(T_{\text{dec}})\pi^2}\right)^{1/4}.
\end{align}
From the equations above, it is clear that large anomaly coefficients are favored from the perspective of entropy dilution, as they enhance the axion decay rate and thereby reduce the amount of entropy injected after the asymmetry is produced.\\

\noindent\textbf{Isocurvature:} 
Since the axion field is already present during inflation and is assumed to be much lighter than the Hubble scale during inflation, $H_I$, its homogeneous initial value is accompanied by inflationary fluctuations. 
These fluctuations have the typical size
\begin{align}
    \delta \theta_i \simeq \frac{H_I}{2\pi v_\phi}\, .
\end{align}
In the parameter region considered here, and assuming that the initial misalignment value is not close to the hilltop of the potential, the baryon asymmetry generated scales linearly with the initial displacement, \(Y_B \propto \theta_i\).
Consequently, fluctuations in \(\theta_i\) lead directly to fluctuations in the baryon abundance,
\begin{align}
    \frac{\delta Y_B}{Y_B} \simeq \frac{\delta\theta_i}{\theta_i}
    \simeq \frac{H_I}{2\pi v_\phi\theta_i}\, .
\end{align}
The corresponding baryon-isocurvature power is
\begin{align}
    \mathcal{P}_{B\gamma}
    \simeq
    \left(\frac{H_I}{2\pi v_\phi\theta_i}\right)^2 .
\end{align}
Since observational limits are commonly quoted for cold-dark-matter (CDM) isocurvature, the baryon isocurvature contribution is mapped onto an effective CDM isocurvature mode with the suppression factor
\begin{align}
    \left(\frac{\Omega_b}{\Omega_{\rm CDM}}\right)^2 \simeq 0.035 .
\end{align}
Using the PLANCK 2018 constraint on uncorrelated CDM isocurvature~\cite{Planck:2018vyg}, this gives the bound
\begin{align}
    \mathcal{P}_{B\gamma} \lesssim 2.4\times 10^{-9}.
\end{align}
Equivalently, the Hubble scale during inflation must satisfy
\begin{align}
    \label{eq:isocurv}
    \frac{H_I}{v_\phi}
    \lesssim
    3\times 10^{-4}\,\theta_i .
\end{align}
Finally, the PLANCK 2018 result constrains the Hubble parameter during inflation to be $H_I\lesssim 6\cdot 10^{13}$ GeV~\cite{Planck:2018jri}, while the most recent joint analysis using BICEP/Keck and PLANCK 2018 data yields a tighter bound of $H_I\lesssim 4.7\cdot 10^{13}$ GeV~\cite{BICEP:2021xfz}.

\subsection{Final asymmetry and viable axion parameters}
\label{subsec:finalasym}
\begin{figure}[h]
    \centering
    {{\includegraphics[width=0.4945\textwidth]{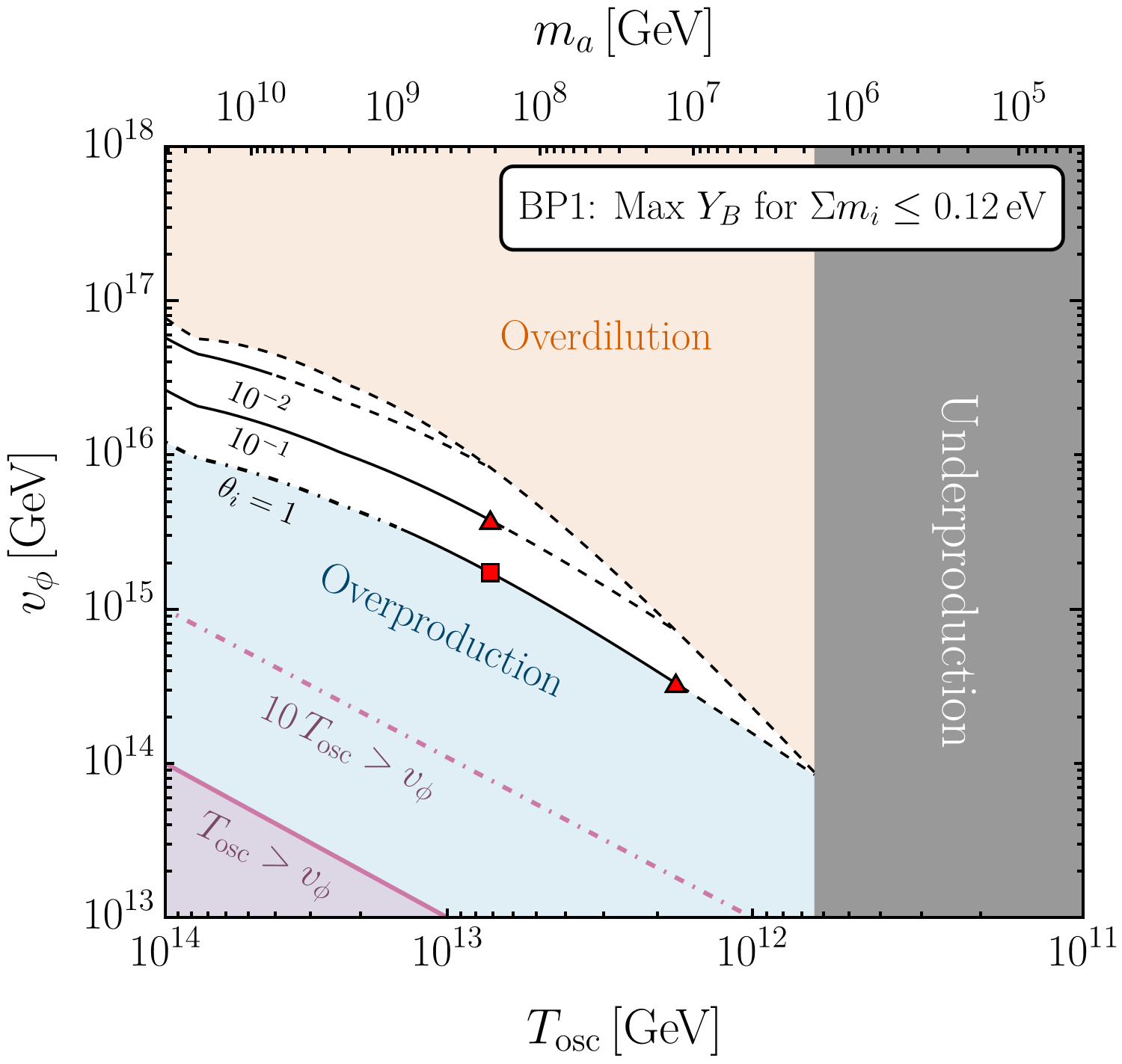}}}
    {{\includegraphics[width=0.4945\textwidth]{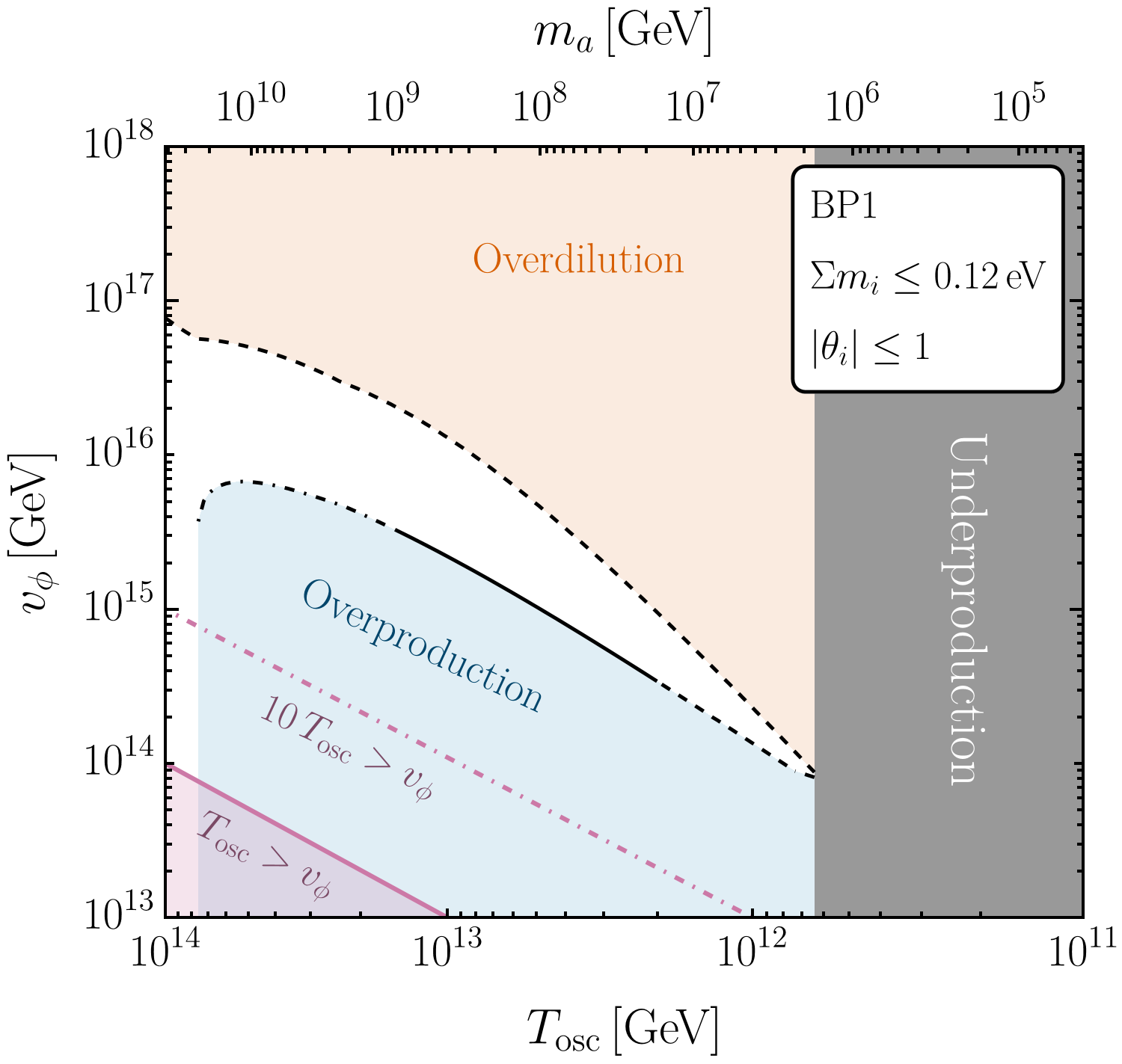}}}
    {{\includegraphics[width=0.4945\textwidth]{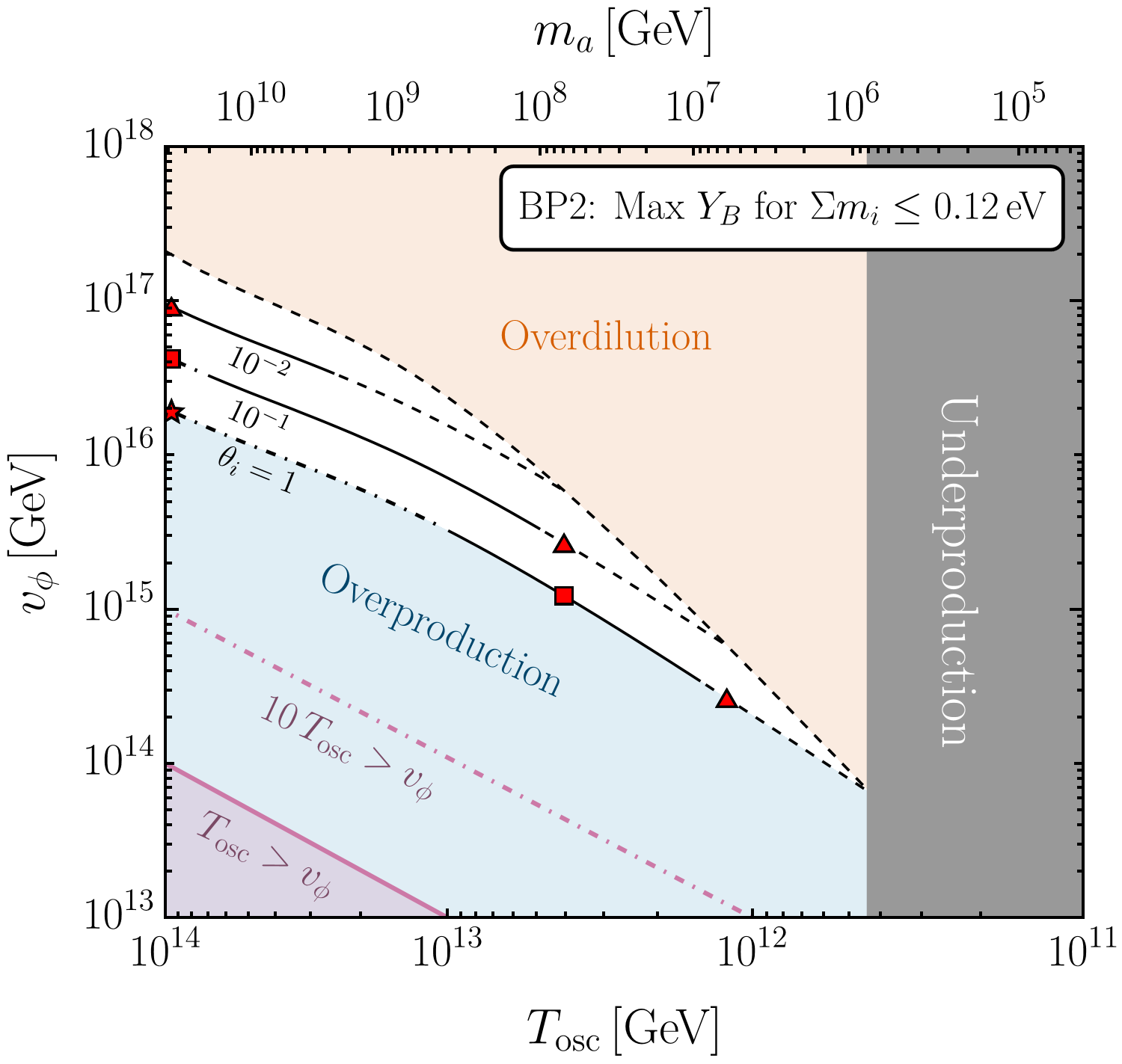}}}
    {{\includegraphics[width=0.4945\textwidth]{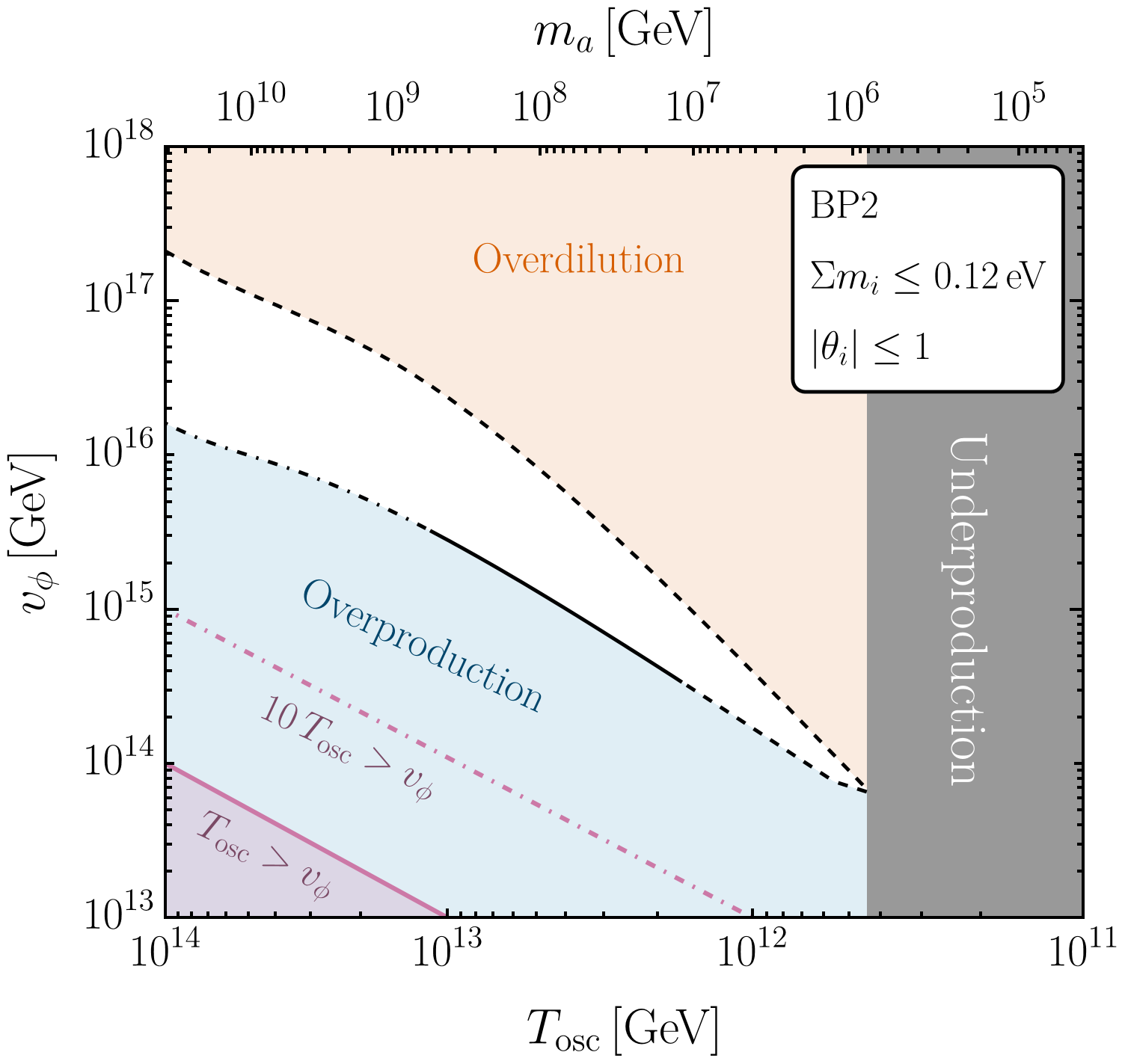}}}
    
    \caption{Panels showing the $T_{\text{osc}}$--$v_\phi$ plane for BP1 (first row) and BP2 (second row). The left panels are obtained from an initial asymmetry corresponding to the upper edge of the blue bands in Fig.~\ref{fig:ParameterBefore}. The right panels show instead the maximal available parameter space for neutrino masses respecting the PLANCK bound and $\theta_i\in (0,1]$. Regions of parameter space where the observed BAU can be successfully reproduced are shaded white, and black bands show solutions for selected fixed values of $\theta_i$ as indicated. The blue region gives baryon overproduction, the orange region gives baryon underproduction due to excessive entropy dilution from axion decays, while in the gray region, the generated baryon asymmetry is below the observed value even in the absence of entropy dilution.   
    Isocurvature constraints are represented by dashed, solid, and dot-dashed lines, which correspond to the intervals $3 \times 10^{-4} \, \theta_i \, (v_\phi / \text{GeV}) \in (10^{10}, 10^{11})$, $(10^{11}, 10^{12})$, and $(10^{12}, 10^{13})$, respectively (see Eq.~\eqref{eq:isocurv}). Triangles, squares, and stars indicate $\Delta_a=0.1$, $\Delta_a=0.01$, and $\Delta_a=0.001$, respectively. See text for discussion.}
    \label{fig:Money}
\end{figure}

Our final results, which encode both isocurvature bounds and entropy dilution from axion decays, are displayed in Fig.~\ref{fig:Money}. The figure illustrates the available parameter space in the $T_{\text{osc}}$--$v_\phi$ plane for BP1 (first row) and BP2 (second row), also showing the corresponding value of $m_a$. The panels to the left are obtained by fixing $m_1$ such that the initial asymmetry matches the upper edge of the blue bands shown in Fig.~\ref{fig:ParameterBefore}. The panels to the right display the parameter space when both the initial misalignment angle $|\theta_i| \in (0, 1]$ and the sum of neutrino masses $\sum m_i \leq 0.12$~eV are treated as free parameters. 

The white regions indicate where the BAU is successfully reproduced for an appropriate choice of $\theta_i \in (0, 1]$,\footnote{Using Eq.~\eqref{eq:Tdec}, one can verify that the axion decays well before Big Bang Nucleosynthesis everywhere within this white region.} whereas the black bands represent contours in the $T_{\text{osc}}$ -- $v_\phi$ plane where selected fixed values of $\theta_i$ yield the correct BAU. The uppermost black band in each plot does not correspond to a fixed $\theta_i$; rather, it is determined by tuning $\theta_i$ such that $Y_{B-L}^{\text{sLG}}(T_{\text{osc}},\,\theta_i)$ is consistent with the BAU, and then extracting the maximal value of $v_\phi$ for which the dilution factor is $\Delta_a \simeq 1$. Additionally, red triangles, squares, and stars on the contour lines denote points where $\Delta_a = 0.1, \, 0.01,$ and $0.001$, respectively. In the gray region to the right in each plot, the initially generated asymmetry $Y_{B-L}^{\text{sLG}}$ is below the observed value. In the orange regions, $v_\phi$ is so large that even if $T_{\text{osc}}$ and $\theta_i$ are chosen such that $Y_{B-L}^{\text{sLG}}(T_{\text{osc}},\,\theta_i)$ matches or exceeds the observed value, subsequent axion decays over-dilute the asymmetry below the observed BAU. Conversely, for any given point in the shaded blue regions, entropy dilution from axion decays is insufficient to reduce the asymmetry to an acceptable level for any choice of $\theta_i\in (0,1]$ at that point in the $T_{\text{osc}}-v_\phi$ plane. Finally, the approximation $v_\phi \gg T_{\text{osc}}$ breaks down near the purple region in the lower-left corners. Isocurvature constraints are encoded via the contour line styles: dashed, solid, and dot-dashed lines correspond to the intervals $3 \times 10^{-4} \, \theta_i \, (v_\phi / \text{GeV}) \in (10^{10}, 10^{11})$, $(10^{11}, 10^{12})$, and $(10^{12}, 10^{13})$, respectively.

The panels in the first column differ in two respects: BP2 accounts for the BAU at lower $T_{\rm osc}$ than BP1, and requires more entropy dilution from axion decays at any given $T_{\rm osc}$. Both observations are explained by the larger initial asymmetry generated in BP2, evident from comparing the blue bands in Fig.~\ref{fig:ParameterBefore}. The large asymmetry in BP2 is due to the doublets with the largest LNV rates, $L_2$ and in particular $L_3$, having sizeable FN charges, so the axion sources the leptonic chemical potentials efficiently, much as a Majoron does in spontaneous leptogenesis when global $U(1)_{B-L}$ is spontaneously broken~\cite{Chun:2023eqc,Kuckenberg:2026oax}. In contrast, the (absolute value of the) FN charges for $L_2$ and in particular $L_3$ are significantly smaller in BP1. 

The second column differs qualitatively for $T_{\rm osc}\gtrsim7\times10^{13}$~GeV, where BP1 has no lower limit on $v_\phi$ from overproduction. The reason is visible in Fig.~\ref{fig:ParameterBefore}: there $Y^{\rm sLG}_{B-L}$ ranges from zero up to $\sim10^{-5}$ for BP1, so no dilution is needed and the bound on $v_\phi$ from overproduction vanishes.

Our results are conservative in several respects. For instance, while the blue regions of the parameter space lead to an overproduction of the baryon asymmetry in our baseline scenario, they could become viable if an independent dilution mechanism is introduced into the cosmological history. Moreover, we have restricted the sum of neutrino masses to $\sum m_i \leq 0.12$~eV, a regime in which the total asymmetry production is relatively insensitive to the exact value of $m_1$ across all values of $T_{\text{osc}}$ for BP2 and $T_{\text{osc}}\lesssim 7\times 10^{13}$ GeV for BP1. This insensitivity is apparent from the thickness of the blue bands in Fig.~\ref{fig:ParameterBefore} and, correspondingly, from the minor differences between the lower edges of the white regions in the left and right panels in each row of Fig.~\ref{fig:Money}. If this cosmological bound on neutrino masses is relaxed, the allowed parameter space can be significantly extended, as shown in Ref.~\cite{Kuckenberg:2026oax}.

In summary, we demonstrated that minimal spontaneous LG from flavor can generate the observed BAU across a large range of values for $T_{\text{osc}}$ and $v_\phi$. By considering two simple benchmark points, we demonstrated that the BAU can be generated for $T_{\text{osc}} \gtrsim 4 \times 10^{11}$~GeV and $v_\phi \in [7\times 10^{13}, 2 \times 10^{17}]$~GeV. We emphasize that these results, which span several orders of magnitude in both $m_a$ and $v_\phi$, were obtained under the minimal assumptions that the axion velocity is sourced by standard misalignment, with initial misalignment angle $\abs{\theta_i}\leq 1$, and that LNV is provided solely by the Weinberg operator. In the presence of dynamical DoF mediating the LNV interactions—such as RHNs in the type-I seesaw mechanism—the scale of LNV decoupling can be lowered, thereby permitting significantly lower values of $T_{\text{osc}}$. Furthermore, relaxing the assumption of standard misalignment to, e.g., kinetic misalignment~\cite{Co:2019jts,Co:2019wyp}, would also open up the possibility of reproducing the BAU at temperatures orders of magnitude below those considered here. This would, in turn, both bring the corresponding axion within the reach of experimental searches and open up the possibility to cogenesis of the BAU and DM~\cite{Co:2020xlh,Co:2020jtv,Barnes:2024jap}.

Finally, as can be seen from Fig.~\ref{fig:Money}, maintaining $T_{\text{rh}} \gg T_{\text{osc}}$ while simultaneously requiring that baryonic isocurvature perturbations remain below the current observational bound necessitates efficient inflationary reheating, as previously pointed out in Refs.~\cite{Kusenko:2014uta,Kuckenberg:2026oax}. In particular, for a given Hubble scale of inflation $H_I$, the upper bound on $T_{\text{rh}}$—corresponding to instantaneous reheating—scales as the geometric mean of $H_I$ and the Planck mass $M_P$,
\begin{align}
    T_{\text{rh}}^{\text{max}}\simeq \left(\frac{90}{\pi^2 g_*(T_{\text{rh}})}\right)^{1/4}\sqrt{M_PH_I}.
\end{align}
This observation motivates a more detailed investigation of spontaneous LG from flavor embedded within an inflationary framework. In such a scenario, the generation of the asymmetry must be tracked continuously from the end of inflation, through the reheating phase, and well into the radiation-dominated epoch. We leave these explorations for future work.

\section{Conclusions}
\label{sec:conclusion}
The FN mechanism provides a dynamical framework for addressing the SM flavor puzzle. If the underlying flavor symmetry is global, its spontaneous breaking yields one or more pseudo-Nambu-Goldstone bosons. In this work, we consider the minimal realization of this setup, which employs a horizontal $U(1)_H$ symmetry whose spontaneous breaking gives rise to an axion-like particle. If $U(1)_H$ is broken before the end of inflation, the axion field acquires an initial misalignment angle $\theta_i$ that is highly homogeneous on cosmological scales. Its subsequent coherent motion toward the minimum of its potential acts as a source of spontaneous $CPT$ violation, which is generically transferred to the SM sector provided that the SM fields carry nonzero FN charges. In the presence of $B-L$ violating processes, this rolling axion can drive the generation of the BAU via the mechanism of spontaneous baryogenesis~\cite{Cohen:1987vi,Cohen:1988kt}.

In this work, we consider the minimal ingredients required to realize spontaneous baryogenesis within a FN framework. To this end, we assume that the axion dynamics are governed by the standard misalignment mechanism. We also adopt a conservative, model-independent approach by parametrizing the necessary $B-L$ violation via the dimension-five Weinberg operator. This framework is robustly motivated by the experimental evidence for nonzero neutrino masses, which suggests LNV NP at high scales. Within its regime of validity, the Weinberg operator captures the leading-order effects of any underlying seesaw mechanism or other UV mechanism associated with neutrino-mass generation. As demonstrated in Appendix~\ref{appendix:FN}, the consistency of the effective Weinberg operator description constrains the FN charges of both the SM lepton doublets and the heavy degrees of freedom introduced in specific UV completions to be small. For this reason, we focused on benchmark scenarios with small FN charge assignments for the lepton doublets, which allowed us to remain agnostic about the heavy states of the underlying UV theory.

After presenting the general Boltzmann equations describing asymmetry generation for generic FN charge assignments in Section~\ref{sec:BEqs}, we analyzed their numerical solutions in Section~\ref{sec:asymgen}. Specifically, we investigated scenarios featuring nonzero FN charges exclusively in either the lepton or quark sectors of the SM, as well as combined scenarios with nonzero charges in both sectors. The FN mechanism generically predicts a large anomaly coefficient for the axion coupling to $G\tilde{G}$, which typically induces efficient asymmetry generation in the quark sector. Moreover, if any lepton doublets are charged under $U(1)_H$, the LNV rates couple directly to the axion motion in the Boltzmann equations—a feature characteristic of a Majoron. However, while a standard Majoron possesses a highly suppressed decay rate that leads to excessive entropy dilution and spoils its viability for minimal spontaneous baryogenesis~\cite{Kuckenberg:2026oax}, the axion considered here features additional couplings that allow it to decay on a much shorter timescale. Consequently, it retains the desirable property of a Majoron—namely, efficient asymmetry production—without the associated drawback of catastrophic entropy dilution. 

Our final results, showing available parameter space in the $T_{\text{osc}}$ -- $v_\phi$ plane compatible with baryonic isocurvature constraints and entropy-dilution from axion decays, were presented in Section~\ref{subsec:finalasym}. There, we showed that 
minimal spontaneous LG from FN can reproduce the BAU across several orders of magnitude in both $m_a$ and $v_\phi$. Specifically, by solving the evolution equations for two benchmark points, we found that the observed BAU can be reproduced for $T_{\rm osc}\gtrsim4\times10^{11}$~GeV and $v_\phi\in[7\times10^{13},\,2\times10^{17}]$~GeV.

Along the way, we have commented on several possible extensions of our work. On the precision side, it could be interesting to embed the framework into a model of inflation and track the asymmetry generation from the end of inflation, through inflationary reheating, and into the radiation-dominated Universe. On the model-building side, it is worth noting that our setup is compatible with supersymmetrization, which could lower its tension with Higgs naturalness, cf. the end of Section~\ref{sec:FN} where we also comment on the strong CP problem. The scale of LNV could also be lowered by introducing e.g. dynamical RHNs, which could make the mechanism viable at even lower values of $T_{\text{osc}}$. Furthermore, increasing the axion velocity by invoking, e.g., kinetic misalignment~\cite{Co:2019jts,Co:2019wyp}, would also open up the possibility of reproducing the BAU at temperatures orders of magnitude below those considered here. Such an extension is appealing as it could make the mechanism both experimentally testable and open up the possibility to cogenesis of the BAU and DM~\cite{Co:2020xlh,Co:2020jtv,Barnes:2024jap}. We leave further exploration of these ideas for future work.\\

\noindent\textbf{Acknowledgments:} We would like to thank Konstantin Kuckenberg, Marco Nardecchia, Davide Racco, Kai Schmitz, Mauro Valli, and Sascha Weber for useful discussions related to this work. A.S. acknowledges support from the research project "SHYNE" (ID: SOE 20240000025), funded by the Italian Ministry of University and Research (MUR) and by the European Union  (NextGenerationEU under the Young Researchers 2024 SoE Action) and support from the ERC STG grant “AstroDarkLS” (grant No. 101117510). Both authors acknowledge the use of Claude (Anthropic) for derivation cross-checks, language editing, and assistance with figure-generation code.

\begin{appendix}

\section{On UV completions of the Weinberg operator and FN charges}
\label{appendix:FN}
In this Appendix, we examine the type-I, type-II, and type-III seesaw models—simple UV completions that generate the Weinberg operator at tree level. The purpose is to demonstrate explicitly how assigning FN charges to the heavy DoF modifies the standard relation between couplings and mass scales in the UV theory and the EFT parametrization $c_{ij}/\Lambda_W$. For each model, we also discuss how FN charge assignments for the heavy DoF and the SM leptons can affect the assumption that the states generating the Weinberg operator remain non-dynamical in the early Universe.  \\

\noindent\textbf{Type-I seesaw:} The type-I seesaw model~\cite{Minkowski:1977sc,Yanagida:1979as,Yanagida:1980xy,Gell-Mann:1979vob,Mohapatra:1979ia, Schechter:1980gr} extends the SM by introducing at least two RHNs $N_\alpha$, with $\alpha$ labeling the RHN species. In the standard implementation without FN charges, one has Yukawa interactions between RHNs and the SM lepton doublets, and a Majorana mass matrix for the RHNs. Upon assigning FN charges to the SM fields and possibly the RHNs, these terms require modification, and take the following form after $U(1)_H$ is spontaneously broken,
\begin{equation}
\mathcal{L}_{\text{type-I}} \supset
-\,\epsilon^{\,n^{\nu}_{\alpha\beta}}\,Y^{\nu}_{\alpha\beta}\,(\bar{L}_\alpha \tilde{H})N_\beta
-\frac{1}{2}\,\epsilon^{\,n^{N}_{\alpha\beta}}\,M_N\, M_{\alpha\beta}\, N^{C}_\alpha N_\beta
+\text{h.c.}
\label{eq:typeI-lag}
\end{equation}
where $n^{\nu}_{\alpha\beta}=|X_{L_\alpha}-X_{N_\beta}|$ and $n^{N}_{\alpha\beta}=|X_{N_\alpha}+X_{N_\beta}|$, $M_N$ denotes the overall RHN mass scale, and we have neglected additional interactions involving the axion. We will employ the standard assumption that the entries in $Y_{\alpha\beta}$ and $M_{\alpha\beta}$ are $\mathcal{O}(1)$ numbers in the following. For $M_N$ around e.g. the GUT scale, we see that (barring accidental cancellations) FN charges of the RHNs cannot be too large without violating the assumption that the lightest RHN mass exceeds both the maximum temperature of the Universe $T_{\rm max}$ and the Hubble scale during inflation $H_I$. In this sense, for $M_N\lesssim M_P$ and large reheating temperatures -- which is a generic requirement to produce a sizable asymmetry through spontaneous LG scenarios based on standard misalignment -- large FN charges in the RHN sector generally necessitate a treatment in which at least some RHNs are dynamical. In such cases, one must also account for effects arising from RHN production and subsequent decays. In conclusion, if the Weinberg operator is UV completed by the type-I seesaw model, the analysis in this paper builds on the implicit assumption that
\begin{align}
    |X_{N_\alpha}+X_{N_\beta}|<\abs{\frac{\log\left(\frac{\text{Max}[T_{\text{max}},\,H_I]}{M_N}\right)}{\log(\epsilon)}},
\end{align}
otherwise, the effective description may break down.

Let us proceed by integrating out the RHNs. It is convenient to absorb the FN
suppression factors into the effective couplings by defining
\begin{equation}
\mathcal{Y}_{\alpha\gamma}\;\equiv\;\epsilon^{\,|X_{L_\alpha}-X_{N_\gamma}|}\,Y^{\nu}_{\alpha\gamma}\,,
\qquad
\mathcal{M}_{\gamma\delta}\;\equiv\;M_N\,\epsilon^{\,|X_{N_\gamma}+X_{N_\delta}|}\,M_{\gamma\delta}\,,
\label{eq:typeI-dressed}
\end{equation}
so that the FN charges of the heavy states enter through the flavor structure of
$\mathcal{M}$ rather than through an overall power of $\epsilon$. In terms of these
quantities, integrating out the RHNs yields
\begin{equation}
\mathcal{L}_{\text{type-I}} \supset
\frac{1}{2}\Big[(L_\alpha \tilde{H})\,
\big(\mathcal{Y}\,\mathcal{M}^{-1}\mathcal{Y}^{T}\big)_{\alpha\beta}\,
(\tilde{H}^{T} L^{C}_\beta)+\text{h.c.}\Big]\,,
\label{eq:typeI-weinberg}
\end{equation}
and comparing with Eqs.~\eqref{eq:FNL}-\eqref{eq:PMNS} we find
\begin{equation}
\frac{c_{\alpha\beta}\,\epsilon^{\,n^{W}_{\alpha\beta}}}{\Lambda_W}
=\frac{1}{2}\big(\mathcal{Y}\,\mathcal{M}^{-1}\mathcal{Y}^{T}\big)_{\alpha\beta}
=\frac{\sum_{i=1}^{3}U^{*}_{\alpha i}U^{*}_{\beta i}m_i}{2v_H^{2}}\,.
\label{eq:typeI-matching}
\end{equation}
Note that the powers of $\epsilon$ cannot in general be pulled out of the matrix
product in Eq.~\eqref{eq:typeI-weinberg}: each term in the implicit sum over the
heavy-flavor indices carries its own FN suppression, and the inversion of
$\mathcal{M}$ mixes different charge assignments whenever $M_{\gamma\delta}$ is
non-diagonal. For diagonal $M_{\gamma\delta}=\delta_{\gamma\delta}$ the expression
simplifies to
\begin{equation}
\frac{c_{\alpha\beta}\,\epsilon^{\,n^{W}_{\alpha\beta}}}{\Lambda_W}
=\frac{1}{2M_N}\sum_{\gamma}
\epsilon^{\,|X_{L_\alpha}-X_{N_\gamma}|-2|X_{N_\gamma}|+|X_{N_\gamma}-X_{L_\beta}|}\,
Y^{\nu}_{\alpha\gamma}Y^{\nu}_{\beta\gamma}\,,
\label{eq:typeI-diagonal}
\end{equation}
and, barring accidental cancellations, the parametric size of
$c_{\alpha\beta}/\Lambda_W$ is controlled by the smallest exponent appearing in
the sum.

In the simplest case, $X_{N_i}=0$, the scale $\Lambda_W$ can be identified with $M_N$,
and $c_{\alpha\beta}$ is, under the standard assumption of anarchic UV couplings, an
$\mathcal{O}(1)$ matrix. Meanwhile, for generic $X_{N_i}$, the coefficient
$c_{\alpha\beta}$ inherits a nontrivial flavor structure governed by the FN charge
assignments, leading to hierarchies and correlations among its entries. Finally, from the second equality in Eq.~\eqref{eq:typeI-matching}, we see that
parametrically $M_N \sim \epsilon^{\,\hat{n}_{\alpha\beta}}\,v_H^{2}/m_\nu$, where
$\epsilon^{\,\hat{n}_{\alpha\beta}}$ denotes the overall FN suppression of
$(\mathcal{Y}\mathcal{M}^{-1}\mathcal{Y}^{T})_{\alpha\beta}$. Hence, even if
$X_{N_i}=0$, all $X_{L_i}$ also need to be small to avoid dynamical RHNs and a
breakdown of the effective description. \\

\noindent\textbf{Type-II seesaw:} The type-II seesaw model~\cite{Magg:1980ut,Lazarides:1980nt,Schechter:1980gr,Mohapatra:1980yp,Cheng:1980qt} extends the SM with an electroweak scalar triplet, $\Delta$, with hypercharge $1$. The Lagrangian includes the following terms, 
\begin{align}
    \mathcal{L}_{\text{type-II}}&\supset -\left[Y^*_{\alpha\beta}\overline{L^C_{\alpha i}}\epsilon_{ij}\Delta_{jk}L_{\beta k}+\text{h.c.}\right]-\left[\mu^* H_i^T\epsilon_{ij}\Delta^\dagger_{jk}H_k+\text{h.c.}\right]-M_\Delta^2\Tr{\Delta\Delta^\dagger}.
\end{align}
where $M_{\Delta}$ and $\mu$ are dimensionful couplings and $Y$ is dimensionless. When this model is embedded into the FN framework, the terms above are modified as follows
\begin{align}
    \mathcal{L}_{\text{type-II}}&\supset -\left[Y^*_{\alpha\beta}\epsilon^{|X_{L_\alpha}+X_{L_\beta}+X_{\Delta}|}\overline{L^C_{\alpha i}}\epsilon_{ij}\Delta_{jk}L_{\beta k}+\text{h.c.}\right]\nonumber \\
    &-\left[\mu^*\epsilon^{|X_{\Delta}|} H_i^T\epsilon_{ij}\Delta^\dagger_{jk}H_k+\text{h.c.}\right]-M_\Delta^2\Tr{\Delta\Delta^\dagger},
\end{align}
where we have assumed $X_{H}=0$.
Integrating out $\Delta$ then yields the Weinberg operator,
\begin{align}
    \mathcal{L}_{\text{type-II}}&\supset
    \frac{\epsilon^{|X_{L_\alpha}+X_{L_\beta}+X_{\Delta}|+|X_{\Delta}|}}{M_{\Delta}^2}\left[\mu Y_{\beta\alpha}(\overline{L_\alpha}\Tilde{H})(\Tilde{H}^T L_\beta^C)+\text{h.c.}\right],
\end{align}
from which we obtain
\begin{align}
    \frac{c_{\alpha\beta}\epsilon^{n_{\alpha\beta}^W}}{\Lambda_W}=\epsilon^{|X_{L_\alpha}+X_{L_\beta}+X_{\Delta}|+|X_{\Delta}|}\times\frac{\mu Y_{\alpha\beta}^T}{M_{\Delta}^2}=\frac{\sum_{i=1}^3 U_{\alpha i}^\ast U_{\beta i}^\ast m_i}{2v_H^2}.
\end{align}
It follows from the triangle inequality that $|X_{L_\alpha}+X_{L_\beta}+X_{\Delta}|+|X_{\Delta}|\geq |X_{L_\alpha}+X_{L_\beta}|$. Hence, it is clear that the validity of the effective description requires both $X_{\Delta}$ and all $X_{L_i}$ to be small in the type-II scenario.\\

\noindent\textbf{Type-III seesaw:} The type-III seesaw model~\cite{Foot:1988aq,Ma:1998dn} extends the field content of the SM by at least two generations of right-handed $SU(2)_L$ triplet fermions, $\Sigma_R$, carrying zero hypercharge. The Lagrangian includes the following terms,
\begin{align}
    \mathcal{L}_{\text{type-III}}\supset -\frac{1}{2}M_\Sigma M_{ij}\Tr{\overline{\Tilde{\Sigma}_R^i}\Sigma_R^j}-(Y_\Sigma)_{\alpha i}\Bar{L}_\alpha\Sigma_R^i\Tilde{H}^T+\text{h.c.},
\end{align}
where $i,j,\alpha$ are generation indices, $\Tilde{\Sigma}_R\equiv C\Bar{\Sigma}_R^T$, $M_\Sigma$ denotes the fermion mass scale, $Y_\Sigma$ and $M_{ij}$ are dimensionless couplings. Embedding the model into the FN framework, one obtains
\begin{align}
    \mathcal{L}_{\text{type-III}}\supset -\frac{1}{2}M_\Sigma\epsilon^{|X_{\Sigma^i}+X_{\Sigma^j}|}M_{ij}\Tr{\overline{\Tilde{\Sigma}_R^i}\Sigma_R^j}-(Y_\Sigma)_{\alpha i}\epsilon^{|X_{L_\alpha}-X_{\Sigma^i}|}\Bar{L}_\alpha\Sigma_R^i\Tilde{H}^T+\text{h.c.},
\end{align}
where by standard assumption, $M_{ij}$ is a dimensionless matrix with $\mathcal{O}(1)$ entries. Upon integrating out the triplet fermions, and defining in analogy with
Eq.~\eqref{eq:typeI-dressed}
\begin{equation}
\mathcal{Y}^{\Sigma}_{\alpha i}\equiv\epsilon^{\,|X_{L_\alpha}-X_{\Sigma_i}|}\,(Y_\Sigma)_{\alpha i}\,,
\qquad
\mathcal{M}^{\Sigma}_{ij}\equiv M_\Sigma\,\epsilon^{\,|X_{\Sigma_i}+X_{\Sigma_j}|}\,M_{ij}\,,
\label{eq:typeIII-dressed}
\end{equation}
we obtain
\begin{equation}
\mathcal{L}_{\text{type-III}} \supset
\frac{1}{2}\Big[(L_\alpha \tilde{H})\,
\big(\mathcal{Y}^{\Sigma}(\mathcal{M}^{\Sigma})^{-1}\mathcal{Y}^{\Sigma T}\big)_{\alpha\beta}\,
(\tilde{H}^{T} L^{C}_\beta)+\text{h.c.}\Big]\,,
\label{eq:typeIII-weinberg}
\end{equation}
from which it follows that
\begin{equation}
\frac{c_{\alpha\beta}\,\epsilon^{\,n^{W}_{\alpha\beta}}}{\Lambda_W}
=\frac{1}{2}\big(\mathcal{Y}^{\Sigma}(\mathcal{M}^{\Sigma})^{-1}\mathcal{Y}^{\Sigma T}\big)_{\alpha\beta}
=\frac{\sum_{i=1}^{3}U^{*}_{\alpha i}U^{*}_{\beta i}m_i}{2v_H^{2}}\,.
\label{eq:typeIII-matching}
\end{equation}
This result is completely analogous to the type-I seesaw case, with
$N_i \to \Sigma_i$, and the remarks made regarding $M_N$ and $X_{N_i}$ apply
verbatim to $M_\Sigma$ and $X_{\Sigma_i}$.

\end{appendix}


\addcontentsline{toc}{section}{References}

\small

\bibliographystyle{JHEP}
\bibliography{arxiv_1}

\end{document}